\documentclass[12pt,preprint]{aastex}
\usepackage{rotating}
\begin{document}
\title{The Formation of Kiloparsec-Scale \ion{H}{1} Holes in Dwarf Galaxies}
\author{Steven R. Warren}
\affil{Department of Astronomy, University of Minnesota, 116 Church St. SE, Minneapolis, MN 55455, USA; 
warren@astro.umn.edu}

\author{Daniel R. Weisz}
\affil{Department of Astronomy, Box 351580, University of Washington, Seattle, WA 98195, USA; dweisz@astro.washington.edu}

\author{Evan D. Skillman}
\affil{Department of Astronomy, University of Minnesota, 116 Church St. SE, Minneapolis, MN 55455, USA; 
skillman@astro.umn.edu}

\author{John M. Cannon}
\affil{Department of Physics and Astronomy, Macalester College, 1600 Grand Avenue, St.
Paul, MN 55125, USA; jcannon@macalester.edu}

\author{Julianne J. Dalcanton}
\affil{Department of Astronomy, Box 351580, University of Washington, Seattle, WA 98195, USA; jd@astro.washington.edu}

\author{Andrew E. Dolphin}
\affil{Raytheon Company, 1151 East Hermans Road, Tucson, AZ 85756, USA; adolphin@raytheon.com}

\author{Robert C. Kennicutt, Jr.}
\affil{Institue of Astronomy, University of Cambridge, Madingley Road, Cambridge CB3 0HA, UK; robk@ast.cam.ac.uk}

\author{B\"{a}rbel Koribalski}
\affil{Australia Telescope National Facility, CSIRO Astronomy and Space Science,
PO Box 76, Epping NSW 1710, Australia; Baerbel.Koribalski@csiro.au}

\author{J\"{u}rgen Ott}
\affil{National Radio Astronomy Observatory, 520 Edgemont Road, Charlotteville, VA 22903, USA; 
jott@nrao.edu }

\author{Adrienne M. Stilp}
\affil{Department of Astronomy, Box 351580, University of Washington, Seattle, WA 98195, USA; adrienne@astro.washington.edu}

\author{Schuyler D. Van Dyk}
\affil{Spitzer Science Center/Caltech, Mailcode 220-6, Pasadena, CA 91125, USA; vandyk@ipac.caltech.edu}

\author{Fabian Walter}
\affil{Max Planck Institut f\"{u}r Astronomie, K\"{o}nigstuhl 17, D-69117 Heidelberg, Germany; walter@mpia.de}

\and

\author{Andrew A. West}
\affil{Department of Astronomy, Boston University, 725 Commonwealth Avenue,
Boston, MA 02215, USA; aawest@bu.edu}

\begin{abstract}

The origin of kpc-scale holes in the atomic hydrogen (\ion{H}{1}) distributions of some nearby dwarf irregular 
galaxies presents an intriguing problem.  Star formation histories (SFHs) derived from resolved 
stars give us the unique opportunity to study past star forming events that may have helped shape the currently visible 
\ion{H}{1} distribution.  Our sample of five nearby dwarf irregular galaxies spans over an order of magnitude 
in both total \ion{H}{1} mass and absolute $B$-band magnitude and is at the low mass end of previously studied systems.  
We use Very Large Array\footnote[1]{The VLA telescope of the National Radio Astronomy Observatory
is operated by Associated Universities, Inc. under a cooperative
agreement with the National Science Foundation.} \ion{H}{1} line data to estimate the energy required to create the centrally 
dominant hole in each galaxy.  
We compare this energy estimate to the past energy released by the underlying stellar populations computed from SFHs derived from 
data taken with the Hubble Space Telescope\footnote[2]{Based on observations made with the NASA/ESA Hubble Space Telescope
which is operated by the Association of Universities for Research in Astronomy, Inc., under NASA contract NAS 5-26555.}.  The 
inferred integrated stellar energy
released within the characteristic ages exceeds our energy estimates for creating the holes 
in all cases, assuming expected efficiencies.  
Therefore, it appears that stellar feedback provides sufficient energy to produce the observed holes.  
However, we find no obvious signature of single star forming events responsible for the observed structures 
when comparing the global SFHs of each galaxy in our sample to each other or to those of dwarf irregular galaxies reported 
in the literature.  We also fail to find evidence of a central star cluster in FUV or 
H$\alpha$ imaging.  We conclude that large \ion{H}{1} holes are likely formed from multiple generations of star formation and only 
under suitable interstellar medium conditions.

\end{abstract}  


\section{Introduction}

Atomic hydrogen (\ion{H}{1}) observations of nearby galaxies reveal complex gas distributions.  In many systems, the neutral
interstellar medium (ISM) contains
holes, shells,  and/or cavities (e.g., \citealt{hei79}; \citealt{bri86}; \citealt{puc92}; \citealt{oey95}; \citealt{kim99}; 
\citealt{walt99}; \citealt{walt01}; \citealt{mul03}; \citealt{rel07}; \citealt{chu08}).  Some holes in dwarf irregular (dIrr) 
galaxies are large enough that they become the dominant feature of the ISM, encompassing most, if not all of the
optical disk (e.g., M81 Dwarf A - \citealt{sar83}, Sagittarius DIG - \citealt{yal97}, Holmberg I - \citealt{ott01}, 
DDO 88 - \citealt{simp05}, DDO 165 - \citealt{can11}).  

It has been suggested that these structures are created by
feedback from stellar processes (e.g., stellar winds and supernovae (SNe) in OB associations; \citealt{wea77}; 
\citealt{cas80}; \citealt{mcc87}; \citealt{ten88}).  Indeed, \citet{ott01}, \citet{simp05}, \citet{wei09}, and \citet{can11} examined the 
stellar content within the \ion{H}{1} holes of Holmberg I, DDO 88, Holmberg II, and DDO 165 respectively, and determined that 
the underlying stellar populations provide sufficient mechanical energy needed to create the observed holes within 
their estimated ages.

However, the premise that stellar winds and SNe are responsible for forming \ion{H}{1} holes has been called into 
question.  
\citet{hei84} pointed out that the largest ($\sim$1 kpc) Galactic ``supershells" seemed to require more energy than is 
available in a typical OB association.  Looking at holes in the Large Magellanic Cloud, \citet{kim99} concluded 
that there is only a 
weak correlation between the locations of H$\alpha$ emission and \ion{H}{1} holes.  \citet{pern04} compared the 
locations of radio pulsars with Galactic holes and concluded that the largest holes are not consistent with the multiple 
SNe from a single-aged cluster formation scenario.  Similarly, \citet{hat05} compared the locations of known \ion{H}{1} holes in the Small Magellanic Cloud 
to OB associations, supergiants, 
Cepheids, Wolf-Rayet stars, SN remnants, and star clusters, and found that there are $\sim$1.5 times as 
many holes without evidence for recent star formation as there are with recent star formation tracers.  Further, the largest 
\ion{H}{1} shell in the Local Group dIrr IC 1613 contains $\sim$27 OB associations in projection (\citealt{bor04}; \citealt{sil06}), 
but the inferred stellar input energy does not account for the estimated hole creation energy \citep{sil06}.  Taken together,
these findings argue against a single-aged cluster being responsible for forming large \ion{H}{1} holes; however, stellar
processes from multiple generations of star formation remains viable.

Other authors have proposed alternative formation hypotheses to the SNe origin.
\citet{efr98} and \citet{loe98} postulate that a high-energy gamma ray burst (GRB) from the death of a single massive star 
could create kpc sized holes in the ISM, thus offering an explanation for holes without a detectable underlying cluster.  These 
authors assume the energy from GRBs is emitted isotropically.  However, GRBs release most of their energy in bi-directional beams
\citep{rho97}, making this scenario less likely to produce large \ion{H}{1} holes. 
Other models of \ion{H}{1} hole formation without SNe are gravitational and thermal instabilities (\citealt{wad00}; \citealt{dib05}), a 
fractal ISM \citep{elm97}, \ion{H}{1} dissolution by UV radiation \citep{vor04a}, and ram pressure stripping \citep{bur02}.

Yet another alternative formation model was suggested by \citet{ten81}, who investigated the effect of high velocity cloud (HVC) 
collisions with the gas disks of galaxies, concluding that 
the amount of energy deposited into the ISM can be of the same order as a SN ($\sim10^{51}$ erg) or more.  One observational 
prediction of this model is a half-circle arc seen in an \ion{H}{1} position-velocity diagram.  The
half-circle arc arises from the gas being pushed to one direction, corresponding to the direction of travel of the HVC.  
Some observational support for this idea is reported by \citet{hei79,hei84} who point out the most energetic Galactic shells 
in their study all have half-circle arc signatures in position-velocity space.  

It is likely that to a certain degree most of the above physical processes must play a role in the formation of at least some 
\ion{H}{1} holes.  To determine which process is the most crucial, \citet{vor05} carried out numerical simulations.
They investigated three of the above \ion{H}{1} hole formation scenarios (multiple SNe, a single GRB, 
and a HVC collision) by simulating the ISM structure of the dIrr galaxy Holmberg I.  
They showed that multiple SNe explosions more accurately reproduce the \ion{H}{1} morphology of Holmberg I than the GRB or HVC models.

The study of \citet{rho99} raised doubt about the stellar feedback from a single-aged cluster idea.  These authors 
investigated the
numerous \ion{H}{1} holes of Holmberg II with Very Large Array (VLA) \ion{H}{1}, ground based $BVR$, and narrowband H$\alpha$ images.  
They expected to find
the remnants of the OB associations (A, F, and G type main sequence stars) left behind in the holes, but instead discovered 
that the integrated light was inconsistent with the required cluster masses needed for a multiple SNe formation scenario from
a single-aged cluster. However, \citet{stew00} found that the most prominent hole in IC 2574 is coincident with a prominent
stellar cluster.
Recently, \citet{wei09} used deep HST photometry to 
resolve some of the stellar population of Holmberg II and showed that all of 
the holes observed with HST contain evidence of multiple stellar generations and are consistent with the holes being formed from the energy
input into the ISM by these past stars.  Their study brought to light the need to reconstruct the star formation histories (and thus 
the available stellar input energy) from the resolved stars using the methodology developed by \citet{dol02}.  
These conclusions have been reinforced by the \citet{can11} investigation of a 
large ($\sim$775 pc) \ion{H}{1} hole in DDO 165.  These authors compared the energy budget of the underlying stellar 
population versus time to estimated hole creation energies.  They conclude that the stars have produced more than enough energy to create the
observed \ion{H}{1} hole.

Evidence is mounting that large \ion{H}{1} holes are not formed by a single-aged stellar cluster but by multiple generations of star
formation working in concert.
In this study we investigate dwarf galaxies in which centrally dominant \ion{H}{1} holes have been identified in order to address the 
multiple stellar generation hypothesis: DDO 181 (UGC 8651), Holmberg I (UGC 5139,
DDO 63, KDG 57) \citep{ott01}, M81 Dwarf A (KDG 52) \citep{sar83}, Sextans A (UGCA 205, DDO 75)  \citep{skil88}, and UGC 
8508 .  Studying the specific problem of large holes in dIrr 
galaxies has some advantages over studies of larger spiral galaxies.  For 
example, dIrr galaxies rotate as solid bodies for the bulk of their \ion{H}{1} disks (e.g., \citealt{skil96}), limiting the effects of 
destructive shearing forces due to differential rotation.  These forces can shorten the lifetimes of the holes in spiral 
arms of larger galaxies.  Also, solid body rotation allows the young stars to remain very near their birth places on 
timescales of $\sim10^{8}$ years (for observational support of this see \citealt{doh02} and for a statistical approach see
\citealt{bast11}).  Given the above, we can use deep HST imaging of regions within the radii of the holes to 
create spatially resolved star formation histories (SFHs).  From the SFHs we calculate a cumulative stellar energy 
budget (i.e., stellar input energy over time). We compare the input stellar energy to the energy required to create the holes 
derived from VLA \ion{H}{1} images.  This comparison gives us a handle on the ability of stellar 
feedback to create large holes in the neutral ISM of dIrr galaxies.  We analyze the SFHs for these five galaxies and
compare them to SFHs of dIrr galaxies without central \ion{H}{1} holes to determine if any obvious hole creation signature 
is seen.  Finally, we search for centralized clusters responsible for forming the hole in each galaxy with FUV and H$\alpha$ imaging.

\section{Data Products \label{observations}}

\subsection{\ion{H}{1} Data}

\ion{H}{1} data were taken from two large atomic hydrogen surveys: The \ion{H}{1} Nearby Galaxy Survey (THINGS; AW0605; \citealt{walt08}) and the
Very Large Array - Advanced Camera for Surveys Nearby Galaxy Survey Treasury (VLA-ANGST; AO0215; \citealt{ott08}).  
Observations 
were made using the VLA in B, C, and D array configurations.  Reduction of the THINGS data has been 
described in detail by \citet{walt08}.  Standard {\sc AIPS} spectral line reduction procedures were followed.  Flux,
bandpass, and phase calibration were performed using VLA calibrators.  The final moment 0 maps were flux corrected
\citep{jor95} and produced images with a natural weighted resolution of $\sim$14$\arcsec$ $=$ $\sim260$ pc for 
Holmberg I and $\sim$15$\arcsec$ $=$ $\sim260$ pc for M81 Dwarf A.

Reduction of the VLA-ANGST data is described in an upcoming paper, but closely followed the procedures of the
THINGS pipeline.  Deviations from the THINGS reduction pipeline were only required where problems arose due to the addition of
Extended VLA (EVLA) antennas into the array.  Baselines between EVLA antennas were affected by power aliased into the first 0.5 MHz
of the baseband in the conversion of the signal from digital to analog.  These baselines were subsequently removed from the data.  
The baselines between VLA and EVLA antennas were unaffected and remain in the data.  The 
final moment 0 maps were flux corrected \citep{jor95} and produced images with a natural weighted resolution of 
$\sim$12$\arcsec$ $=$ $\sim175$ pc for DDO 181, $\sim$11$\arcsec$ $=$ $\sim75$ pc for Sextans A, and $\sim$13$\arcsec$ $=$ 
$\sim165$ pc for UGC 8508.  We show the \ion{H}{1} integrated intensity maps for the sample galaxies in the top panels of 
Figures \ref{ddo181_hole} - \ref{u8508_hole}.  Table \ref{imageprops} gives the relevant \ion{H}{1} image properties for each 
galaxy and Table \ref{genprops} gives the general observed properties of each galaxy in our sample.

\subsection{HST Data}

HST/ACS and WFPC2 imaging were obtained as part of three HST programs.  Data for Sextans A were
first described in \citet{doh97} and \citet{doh02} and were reprocessed by \citet{holt06}.  Imaging of Holmberg I and M81 Dwarf A were
obtained as part of a larger HST program aimed at studying M81 Group dwarf galaxies (GO-10605; PI: Skillman)\citep{wei08} and were 
reprocessed along with DDO 181 and UGC 8508 within the ACS Nearby Galaxy Survey Treasury (ANGST) project \citep{dal09}.
Here, we briefly summarize the details of the photometry and measurements of the star formation histories.  Full details of the
photometry are listed in \citet{holt06} and \citet{dal09}, while details of the star formation histories are listed in 
\citet{wei08} and \citet{wei11}.  
Following standard data reduction with the HST pipeline, photometry for WFPC2 data was performed using HSTphot \citep{dol00} 
while the ACS data were processed with the ACS-specific module of DOLPHOT \citep{dol00}.
The raw photometry was then filtered to exclude non-stellar point spread functions, 
and the resultant photometric catalogs range from a low of 17,450 stars for M81 Dwarf A to 121,198 stars for 
Holmberg I.  Approximately 500,000 artificial star tests were performed for each galaxy to compute the completeness 
functions and quantify uncertainties due to observational effects. The bottom panels of Figures \ref{ddo181_hole} - 
\ref{u8508_hole} show the F814W images of each galaxy.

\subsection{FUV and H$\alpha$ Data}

We utilize FUV and H$\alpha$ imaging obtained through the 11 Mpc H$\alpha$ UV Galaxy Survey (11HUGS; \citealt{ken08}).  Full details
of the observations and calibrations can be found in \citet{ken08} and \citet{lee11}.  Briefly, standard GALEX calibration and data 
processing were used following procedures outlined in \citet{mor07}.  The GALEX UV images were initially cleaned of 
foreground stars and background galaxies before photometry was performed.  Background levels were estimated from pixels beyond the 
$B$-band 25 mag arcsec$^{-1}$ isophote where UV galaxy flux is assumed to be zero. The background flux is required to estimate the
asymptotic magnitude of the galaxy.  Azimuthally averaged surface brightness profiles were made and a cumulative magnitude versus 
cumulative-magnitude gradient was fit by a line, where the y-intercept is determined to be the asymptotic magnitude of the galaxy.

H$\alpha$ imaging was obtained from three different telescopes between 2001-2005: the Steward Observatory Bok 2.3m telescope on Kitt Peak, 
the Lennon 1.8m Vatican Advanced Technology Telescope (VATT) on Mt. Graham, AZ, and the 0.9m telescope at the Cerro Tololo Interamerican 
Observatory (CTIO).  Common {\sc IRAF} reduction procedures were followed.  $R$-band imaging was scaled to the continuum level of 
narrowband imaging containing H$\alpha$ and [\ion{N}{2}]$\lambda\lambda$6548,84 emission lines.  The H$\alpha$ + [\ion{N}{2}] images were 
then continuum subtracted to isolate the emission lines.  Finally, images were flux calibrated using spectrophotometric standard stars.

\section{Hole Definitions\label{holedef}}

\begin{figure}
\begin{center}
\includegraphics[width=58mm]{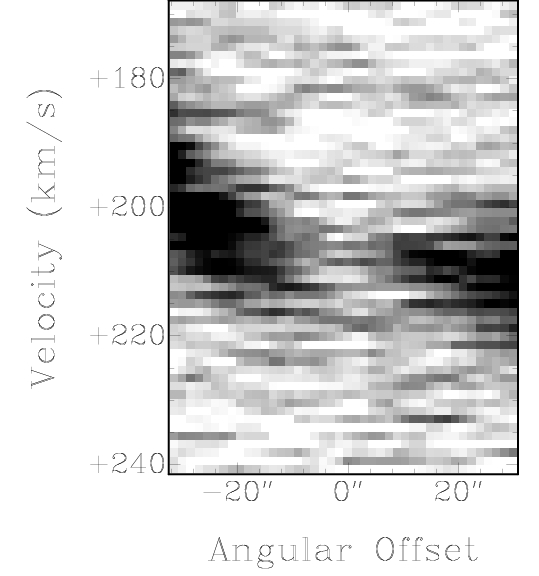}   \includegraphics[width=68mm]{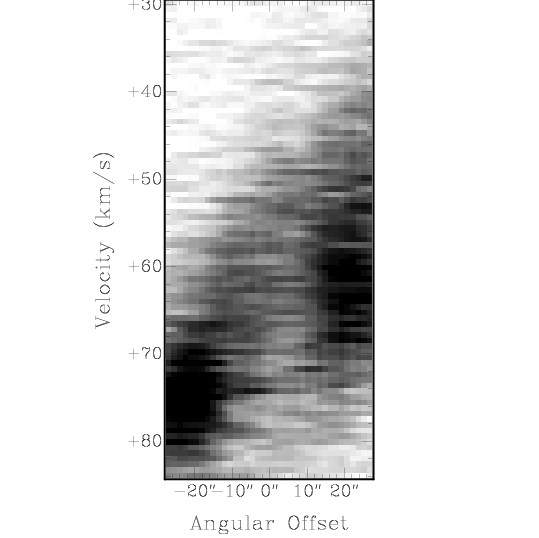}\\
\includegraphics[width=58mm]{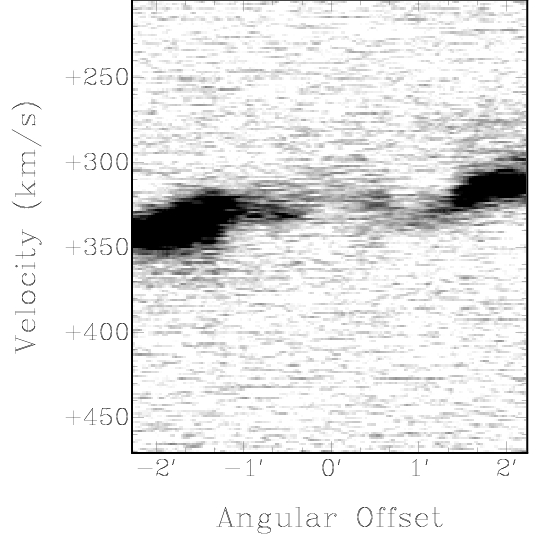} \includegraphics[width=58mm]{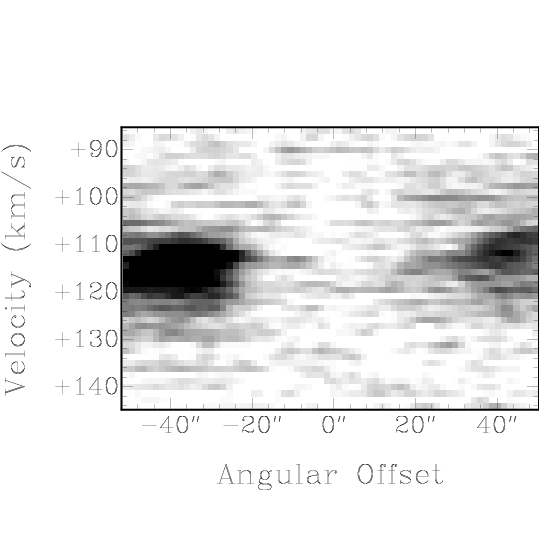} \\
\includegraphics[width=68mm, angle=-90]{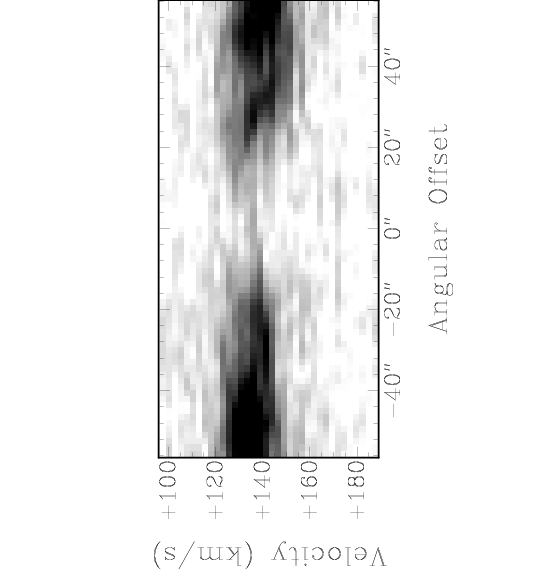}  \\
\end{center}
\caption{Position-velocity diagrams for the centrally dominant holes in (top to bottom, left to right) DDO 181, UGC 8508, 
Sextans A, M81 Dwarf A, and Holmberg I. 
Each diagram is discontinuous, representing a type 1 hole.  Expansion velocities for these galaxies can not be measured 
from these diagrams since they are indistinguishable from the average velocity dispersion in the galaxy.  The grey scale 
reflects the \ion{H}{1} intensity.
\label{pv}}
\end{figure}

Holes in the \ion{H}{1} distribution of dIrr galaxies come in a variety of shapes and sizes.  \citet{bri86} defined three different types
of holes depending on their characteristics in a position-velocity (P-V) diagram.  A Type 1 hole is a structure 
that has expanded out of the disk (i.e., blow-out) and shows a discontinuous profile in a P-V diagram.  Type 1 holes are among the 
most commonly observed as they are readily visible as depressions in \ion{H}{1} integrated intensity maps.  Type 2 holes are those 
offset from the plane of the host galaxy such that only one side has undergone blow-out.  These holes have a half-circle arc 
signature in a P-V diagram similar to what might be observed from a HVC collision (e.g., \citealt{vor05}).  Finally, a Type 3 hole 
is a ``classic" hole where a coherent expanding structure can be seen in a P-V or radius-velocity diagram.  This hole type allows for 
its current expansion velocity to be 
measured directly.  Figure \ref{pv} shows a representative P-V diagram of each galaxy.  All of the centrally dominant holes in 
this study are Type 1 holes, that is, their P-V diagrams are discontinuous and show no signs of expansion discernible from their
average velocity dispersions.  We are therefore unable to define expansion velocities for the holes and instead use the average velocity 
dispersion as a proxy for this value; we discuss this further in \S\ref{time}.

\begin{figure}
\begin{center}
\includegraphics[width=120mm]{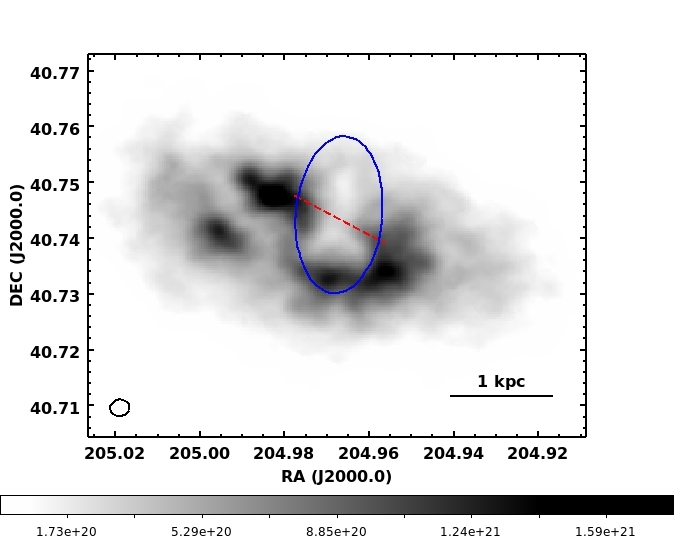} \\
\includegraphics[width=140mm]{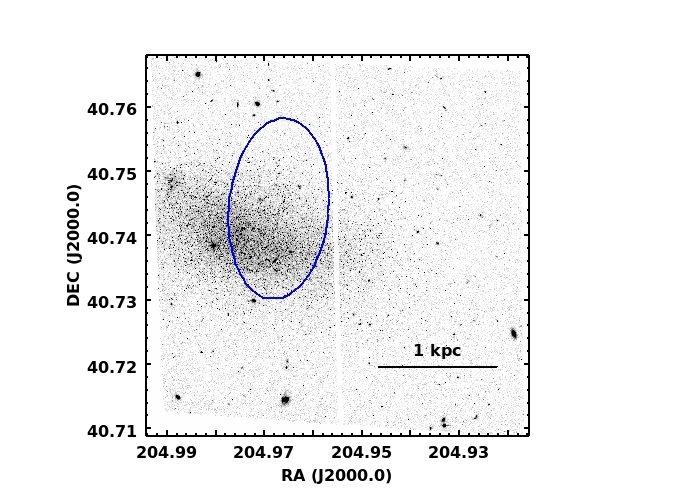} \\
\end{center}
\caption{VLA \ion{H}{1} integrated intensity map (top) and F814W HST image (bottom) of DDO 181.
The blue ellipses denote the area of the hole.  The red dashed line shows the cut used
to make the P-V diagram for Figure \ref{pv}.  The units of the scale bar for the \ion{H}{1} image are cm$^{-2}$ and the black
ellipse at the lower left denotes the size and shape of the synthesized beam.
\label{ddo181_hole}}
\end{figure}

\begin{figure}
\begin{center}
\includegraphics[width=110mm]{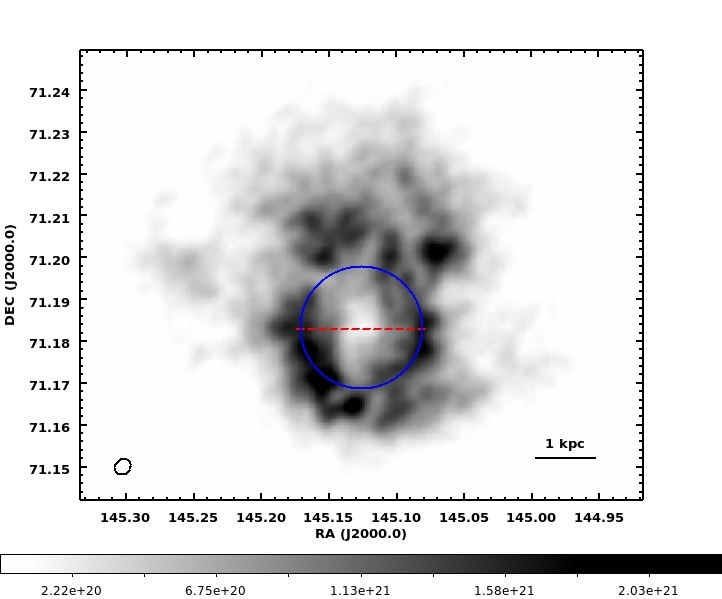} \\
\includegraphics[width=120mm]{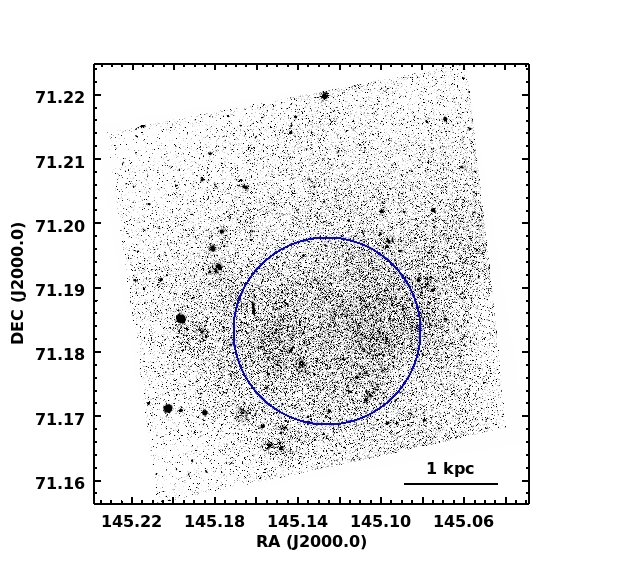} \\
\end{center}
\caption{VLA \ion{H}{1} integrated intensity map (top) and F814W HST image (bottom) of Holmberg I. The blue circles denote 
the area of the hole.  The red dashed line shows the cut used
to make the P-V diagram for Figure \ref{pv}.  The units of the scale bar for the \ion{H}{1} image are cm$^{-2}$ and the black
ellipse at the lower left denotes the size and shape of the synthesized beam.\label{hoi_hole}}
\end{figure}

\begin{figure}
\begin{center}
\includegraphics[width=115mm]{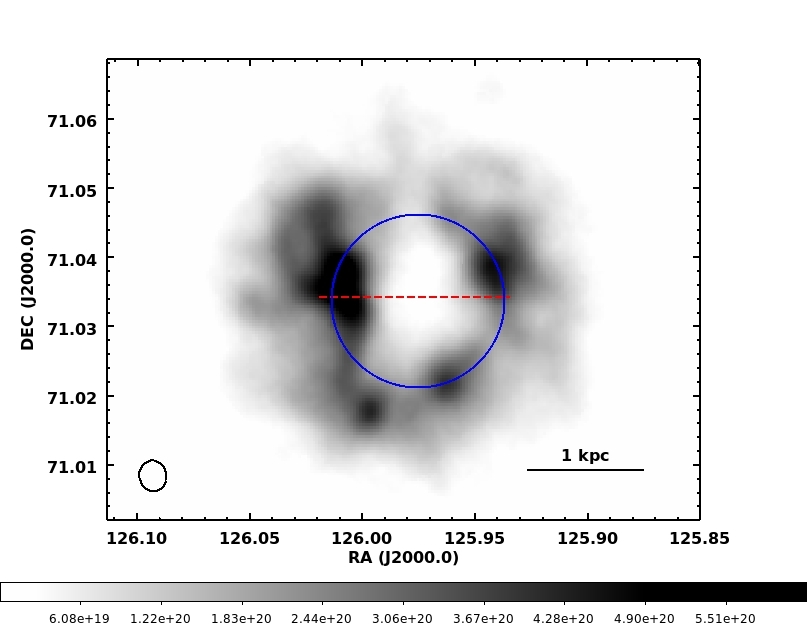} \\
\includegraphics[width=123mm]{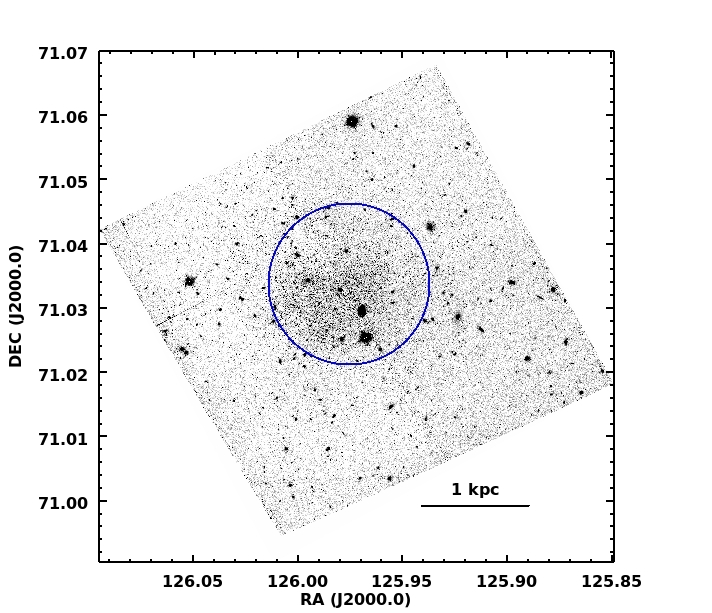} \\
\end{center}
\caption{VLA \ion{H}{1} integrated intensity map (top) and F814W HST image (bottom) of M81 Dwarf A. The blue circles denote the 
area of the hole.  The red dashed line shows the cut used
to make the P-V diagram for Figure \ref{pv}.  The units of the scale bar for the \ion{H}{1} image are cm$^{-2}$ and the black
ellipse at the lower left denotes the size and shape of the synthesized beam.\label{m81da_hole}}
\end{figure}

\begin{figure}
\begin{center}
\includegraphics[width=103mm]{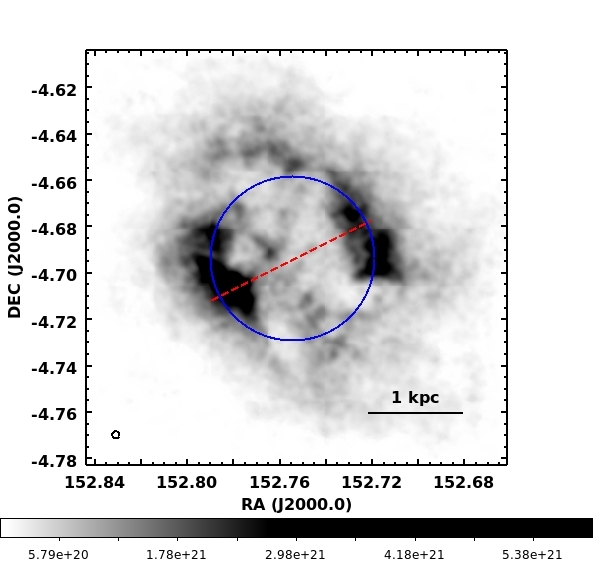} \\
\includegraphics[width=105mm]{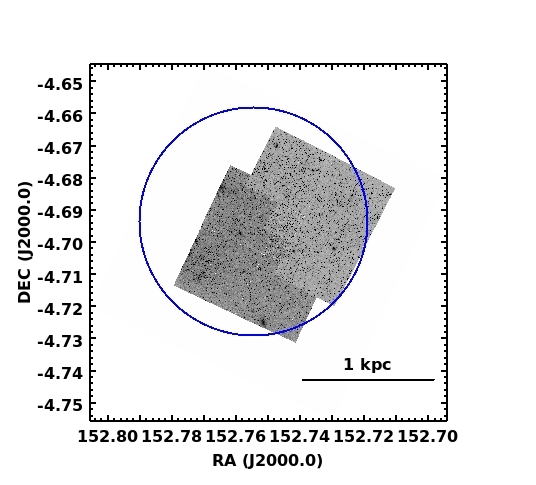} \\
\end{center}
\caption{VLA \ion{H}{1} integrated intensity map (top) and F814W HST image (bottom) of Sextans A. The blue circles denote 
the area of the hole.  The red dashed line shows the cut used
to make the P-V diagram for Figure \ref{pv}.  The units of the scale bar for the \ion{H}{1} image are cm$^{-2}$ and the black
ellipse at the lower left denotes the size and shape of the synthesized beam.\label{sexa_hole}}
\end{figure}

\begin{figure}
\begin{center}
\includegraphics[width=115mm]{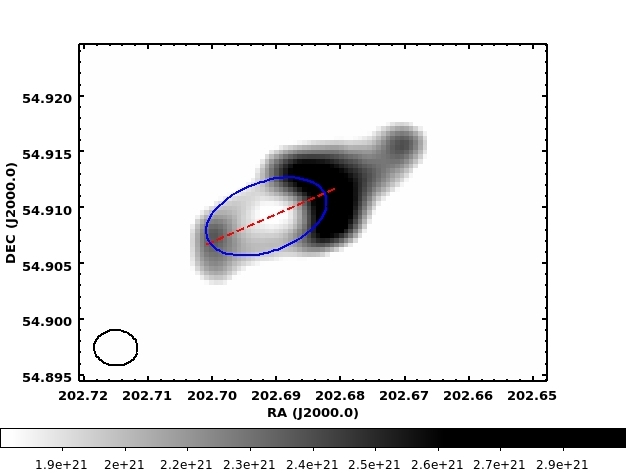} \\
\includegraphics[width=130mm]{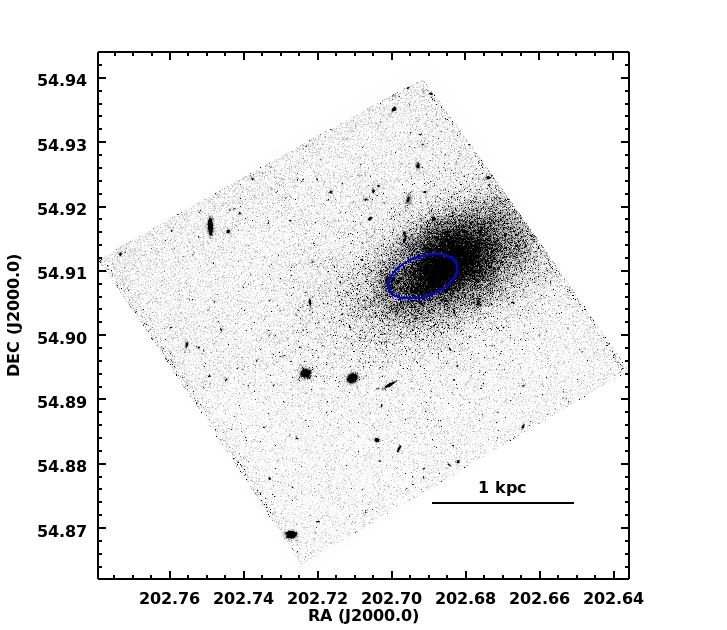} \\
\end{center}
\caption{VLA \ion{H}{1} integrated intensity map (top) and F814W HST image (bottom) of UGC 8508. The blue ellipses denote the area of 
the hole.  The red dashed line shows the cut used
to make the P-V diagram for Figure \ref{pv}.  The units of the scale bar for the \ion{H}{1} image are cm$^{-2}$ and the black
ellipse at the lower left denotes the size and shape of the synthesized beam.\label{u8508_hole}}
\end{figure}

We define the locations and shapes of the holes visually from the \ion{H}{1} integrated intensity maps.  We assume circular shapes
for Holmberg I, M81 Dwarf A, and Sextans A, but define elliptical shapes for the holes in DDO 181 and UGC 8508.  Defining holes
visually from the integrated intensity map is straightforward for the large structures in our sample of galaxies.  Recent work by
\citet{bag11} also identified holes visually on a sample of 20 THINGS galaxies, although they also used P-V diagrams in order to 
identify smaller, less obvious holes.  Automated search algorithms designed to detect \ion{H}{1} holes do exist (e.g., \citealt{thil98};
\citealt{mash99}; \citealt{daig03}; \citealt{ehle05}).  These routines are used mainly to find holes not readily seen in the
integrated intensity maps but within the spectral data cubes.  Since our holes are all easily identifiable in the integrated intensity
maps (Figures \ref{ddo181_hole} - \ref{u8508_hole}) these algorithms are unnecessary for this work. 

The large hole identified in
each galaxy is a single structure except for that of Sextans A.  Sextans A has a complex \ion{H}{1} distribution consisting of a general
depression surrounded by two larger, higher column density clouds.  The central depression seems to be filled with multiple holes
of various sizes and shapes \citep{puc94}.  However, previous authors, based on lower angular resolution data  
(e.g., \citealt{skil88}; \citealt{vand98}) have described this feature as a single hole.  This gives us a unique opportunity to
test the effect of resolution on our observed holes.  The holes we observe in the more distant galaxies of our sample may also be
filled with numerous, smaller holes that we are not sensitive to.  If we describe the \ion{H}{1} morphology of Sextans A as a
single, large hole the amount of energy we derive for its creation will be larger than required to form many smaller holes.  So if
it is shown that the underlying stars have ample energy to create one large hole, then certainly they could produce many smaller
holes; thus we include it here and treat it as a single, central hole.

\begin{figure}
\includegraphics[width=168mm]{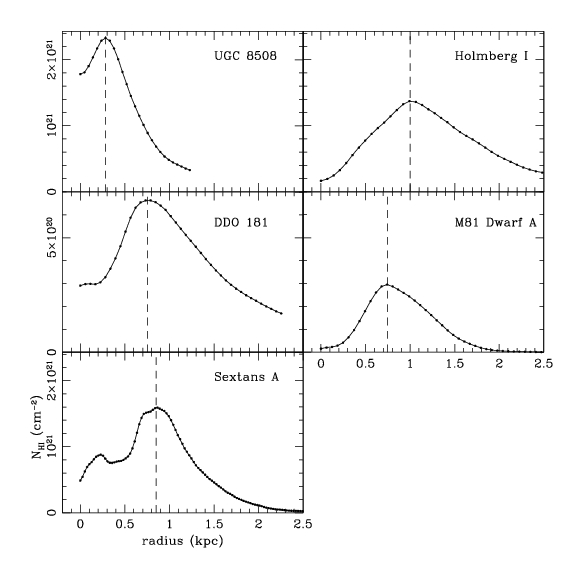}
\caption{Azimuthally averaged column density versus radius for each galaxy.  The vertical dashed lines denote the adopted radii of 
the holes. \label{rp}}
\end{figure}

The radial extents of the holes were defined by plotting the average \ion{H}{1} column density versus radius as was done previously
by \citet{ott01} and \citet{simp05}.  This procedure was done 
with the {\sc IRAF} task {\sc ELLIPSE} which computes the average value on the boundary of an input ellipse incrementally from the starting 
location outward in predefined step sizes.  The hole radius is then defined as the peak in this radial profile.  If a hole is
created in a uniformly dense medium, the hole boundary will be defined as the location of the shock front and will be identified as
a delta function in a similar radial profile.  Once the hole stops expanding, the shock front will broaden and disperse over time.
 We do not find any evidence for a shock front, but if the gas were being piled up
our assumption of treating the peak of this profile as the radius of the hole is reasonable. Figure 
\ref{rp} shows the radius versus average column density profiles.  For the elliptical holes in DDO 181 and UGC 8508 the radius is defined 
as the semi-major axis length.  Each profile rises to a maximum before declining again, except for the \ion{H}{1} profile of Sextans A.  
The radial profile for Sextans A shows substructure before the global peak.  This substructure is due to the non-uniform nature of the central
depression mentioned above, including a region of higher column density gas clumped just offset from the defined hole center.  
The holes range in radius from $\sim$285 pc (UGC 8508) to $\sim$970 pc (Holmberg I).  We plot the hole boundary with a blue circle 
or ellipse on the \ion{H}{1} and optical images (Figures \ref{ddo181_hole} - \ref{u8508_hole}) as well as list their individual 
properties in Table \ref{holeen}.

\section{Energetics \label{energy}}
 
We use a blast wave model to estimate the total amount of energy needed to make a cavity in the ISM.  We then compare this
to an estimate of the stellar feedback energy derived from the resolved stellar population to test the idea that the energy from 
stellar feedback is the dominant source in shaping the neutral ISM in these galaxies.  To make this comparison we determine three 
things: 1) the amount of energy that was required to form the \ion{H}{1} holes, 2) the amount of available energy from stellar 
feedback, and 3) the relevant timescale for energy injection. A timescale in which to compare these 
two calculations is not so straightforward and is described below.

\subsection{Timescales \label{time}}

We must first determine the
relevant timescale over which the input stellar energy shaped the hole.  One such timescale is the kinematic age of the hole,
defined as the ratio of the radius of the hole by the expansion 
velocity.  Looking
at different cuts in a position-velocity diagram we find no evidence for current expansion in any of these holes.  This behavior is 
expected for holes that
have blown out of the disk as the input energy will preferentially follow the path of least resistance (along the blow out
direction).  Further expansion will quickly stall out and become indistinguishable from the velocity dispersion of the
background gas distribution (e.g., \citealt{vor04b}).  

Since
there is no evidence that the holes are currently expanding, we use the average velocity dispersion of the gas as a proxy for
the expansion velocity.  We define the average velocity dispersion as the mean value of the second moment map over the entire galaxy.  
For holes that have not blown out of the disk, using this value
will provide an upper limit to the true age of the hole, because it assumes the expansion rate was constant throughout the
evolution, which certainly was not true during the initial formation stages.  However, since all of our holes are of Type 1,
the kinematic age can be thought of as a characteristic age ($\tau_{char}$), not as an upper or lower limit, since we can not 
know for sure how long
the structure has maintained its current configuration.  Our computed characteristic ages are all roughly 100 Myr (see Table
\ref{holeen})
with the exception of UGC 8508 which has an age $\sim$30 Myr.  These ages fall in the range of
reported hole ages in other galaxies of $\sim$10$^{6}$-10$^{8}$ yr (e.g., \citealt{oey95}; \citealt{walt99}; \citealt{hat05}; 
\citealt{wei09}; \citealt{bag11}).

\subsection{Energy required to form \ion{H}{1} holes \label{HIen}}

In 1974, \citeauthor{che74} derived an empirical relation between the total energy of an explosion, $E_{Hole}$, the 
radius, $r$, and the expansion velocity, $v$, of the cavity left behind, taking into account the initial volume density, $n_{o}$,
of the medium:
\begin{equation} \label{ehole}
E_{Hole}~=~5.3\times10^{43}~\left(\frac{n_{o}}{\mathrm{cm^{-3}}}\right)^{1.12}~\left(\frac{r}{\mathrm{pc}}\right)^{3.12}~\left(\frac{v}
{\mathrm{km~s}^{-1}}\right)^{1.40}~\mathrm{erg}.
\end{equation}
This equation has been used by many authors to estimate the energy required to make \ion{H}{1} holes (e.g.,
\citealt{ott01}; \citealt{simp05}; \citealt{wei09}). However, one must keep in mind the many assumptions that go into using
Equation \ref{ehole}.  First, Equation \ref{ehole} assumes a homogeneous medium throughout the entire evolution of the blast 
wave. For smaller holes that have not blown out of the disk, this assumption provides an {\it upper} limit to the required 
energy, since a realistic ISM density profile is stratified, allowing the shock wave to propagate 
with less resistance and grow to larger sizes.  In the scenario of larger holes that have blown out of the disk, Equation 
\ref{ehole} just gives us an estimate
of the needed energy, but can not be interpreted as an upper or lower limit.  Secondly, defining the initial volume 
density for the cavity is non-trivial.  One needs to know the thickness (column density) and distribution 
(e.g., exponential, Gaussian, etc.) of the gas to accurately define the volume
density at any given point of an \ion{H}{1} map.  Further complicating matters, $n_{o}$ is the $initial$ 
gas volume density and since there already exists an evacuated cavity, this number can only be estimated.  
The uncertainty in the calculated energy value may be as high as an order of magnitude and possibly more depending 
on the true initial conditions.  For example, if one assumes a value for $n_{o}$ of 0.01 whereas the `true' value should be 0.1,
the derived energy value will be off by $\sim$8\%, assuming the radius and expansion velocity used are correct.

To address these issues, we have decided to follow the method of \citet{ott01} to estimate a lower limit to the volume 
density.  For this we assume the gas follows a Gaussian distribution from the midplane, which leads to 
\begin{equation} \label{colden}
N_{\mathrm{HI}}~=~\sqrt{2\pi}hn_{o}
\end{equation}
where $N_{\mathrm{HI}}$ is the \ion{H}{1} column density, $h$ is the gas scale height, and $n_{o}$ is the midplane gas volume
density.  Thus in principle, a measurement of $h$ leads to an estimate of $n_{o}$.  Defining the gas scale height, however, 
is not straightforward.  

There have been many ideas put forth as to 
how one would correctly estimate the gas scale height from the observables of face on galaxies (e.g., \citealt{vdk81}; 
\citealt{pad01}; \citealt{elm01}; \citealt{wei09}).  \citet{wei09} and \citet{can11} used a method described in 
\citet{oh08} to define the scale height as a function of radius.  This process involves defining the stellar mass surface
density of the galactic disk from Spitzer 3.6 $\mu$m imaging, and then assuming that the stars are the only significant source of 
surface density.  For galaxies with baryonic masses greater than 
$\sim$10$^{\mathrm{9}}$ M$_{\mathrm{\sun}}$, the stars in the disk should dominate the local gravitational potential 
(see \citealt{mac99} and references therein).  However, for less massive galaxies,
the dark matter plays an increasing role in the gravitational potential and the gas scale height will be 
smaller than that derived from the stellar distribution alone.  \citet{bane11} recently modeled the effects of dark matter,
gas self-gravity, and the stellar content of four dwarf galaxies, including Holmberg II, on the H I vertical scale height.  Their
derived scale height as a function of radius is generally smaller than the one derived by \citet{wei09}.  The effect is more
pronounced at larger radii.  Unfortunately, we do not have the correct inputs to follow the method of \citet{bane11}, thus for 
uniformity we follow the steps of \citet{ott01} who used the gas scale
height equation derived by \citet{vdk81} to relate the gas scale height to the average gas velocity dispersion, $\sigma_{\mathrm{gas}}$,
and the disk total mass density, $\rho_{\mathrm{t}}$:
\begin{equation} \label{scheight}
h(R)~=~\frac{\sigma_{\mathrm{gas}}}{\sqrt{4 \pi G \rho_{\mathrm{t}}}}.
\end{equation} 
The gas velocity dispersion should be affected by the underlying dark matter potential as well as the stellar mass distribution.

\citet{ott01} combined Equations \ref{colden} and \ref{scheight} yielding
\begin{eqnarray} \label{newsh}
h~&=&~(\sqrt{8\pi}Gm_{p})^{-1}\frac{\sigma_{\mathrm{gas}}^{2}}{(\rho_{\mathrm{t}}/\rho_{\mathrm{HI}})N_{\mathrm{HI}}}
  \nonumber \\
  &=&~579\left(\frac{\sigma_{\mathrm{gas}}}{10~\mathrm{km~s}^{-1}}\right)^{2}\left(\frac{N_{\mathrm{HI}}}
  {10^{21}~\mathrm{cm}^{-2}}\right)^{-1}\left(\frac{\rho_{\mathrm{HI}}}{\rho_{\mathrm{t}}}\right)~\mathrm{pc,}
\end{eqnarray}
where $G$ is the gravitational constant, $m_{p}$ is the proton mass, and $\rho_{\mathrm{HI}}$ is the \ion{H}{1} mass density.  
Similarly to \citet{ott01}, we assume that ($\rho_{\mathrm{HI}}/
\rho_{\mathrm{t}}$) = (M$_{\mathrm{HI}}$/M$_{\mathrm{t}}$) with M$_{\mathrm{t}}$ = M$_{\mathrm{HI}}$ + M$_{\mathrm{He}}$ +
M$_{\mathrm{stars}}$, M$_{\mathrm{He}}$ = 0.3M$_{\mathrm{HI}}$, and a stellar mass-to-light ratio 
(M$_{\mathrm{stars}}$/L$_{B}$) = 1.
We then use the M$_{\mathrm{HI}}$, L$_{B}$ and the average N$_{HI}$ values given in Table \ref{genprops} to derive
($\rho_{\mathrm{HI}}/\rho_{\mathrm{t}}$), $h$, and midplane $n_{o}$.  Finally, we calculate
{\it lower} limits to the input energies (E$_{min}$) required to create the hole in the \ion{H}{1} distribution by substituting
Equations \ref{colden} and \ref{newsh} into Equation \ref{ehole}.

\citet{skil87} find that massive stars form around \ion{H}{1} column density peaks of $\gtrsim$~$10^{21}$ 
cm$^{-2}$.  Therefore, we can derive $upper$ limits to the required input energy (E$_{max}$) for hole
formation by assuming this $N_{\mathrm{HI}}$ value for each galaxy.  
This new energy value represents an upper limit because using $N_{\mathrm{HI}}~=~10^{21}$ cm$^{-2}$ assumes the entire ISM within 
the radius of the hole was at this value prior to forming the hole. This was not the case for each of the galaxies in
our sample, with the possible exception of UGC 8508.  The column density of the gas within and around the hole in UGC 8508 is on 
the order of $10^{21}$ 
cm$^{-2}$; thus the resulting energy estimate may be closer to the expected value than it is for the holes in the other galaxies.
The new $N_{\mathrm{HI}}$ value leads to new estimates of $n_{o}$ using Equations \ref{newsh} and
\ref{colden}.  Equation \ref{ehole} then leads us to upper limits 
to the amount of hole formation energy.  Table \ref{holeen} gives the estimated lower and upper limits required to form the holes 
for each galaxy.  

An alternative method of calculating hole formation energies to the \citet{che74} model was proposed by \citet{mcc87} and implemented 
recently by \citet{cha11}.  The \citet{mcc87} model disregards the single stellar explosion model in favor of continuous energy injection.  
They propose the number of supernovae required to make the hole is given by
\begin{equation} \label{alten}
(N_{*}E_{51}/n_{0})~=~(R_{S}/97~\mathrm{pc})^{2}(V_{S}/5.7~\mathrm{km~s}^{-1})^{3},
\end{equation}
where N$_{*}$ is the number of supernovae, E$_{51}$ is the energy per supernova in units of 10$^{51}$ erg, $n_{0}$ is the initial volume 
density of the gas, R$_{S}$ is the radius of the hole, and V$_{S}$ is the current expansion velocity of the hole.  This equation provides 
values slightly lower than those given by Equation \ref{ehole}.  For example, using the radius and velocity dispersion defined
above for DDO 181 in place of R$_{S}$ and V$_{S}$, respectively, Equation \ref{alten} gives us 105 supernovae = 
1.05$\times$10$^{53}$ erg. This value is lower than the 5.3$\times$10$^{53}$ erg that Equation \ref{ehole} provides us with.  
The larger values provided by Equation \ref{ehole} give us more confidence that we are truly deriving upper limits for the amount of 
energy required to make these large holes.  The lower limits that we derive using Equation \ref{ehole} are also slightly higher
than what Equation \ref{alten} would provide for the same reasons as the above example.  However, if the amount of available energy 
from the stars (see \S\ref{staren}) far exceeds our upper limit derived using Equation \ref{ehole}, the value used for our lower 
limit is less important.  Our estimates for the amount of energy needed to create the observed \ion{H}{1} holes are in the ranges 
derived by other authors for similar holes in other galaxies (e.g., \citealt{ott01}; \citealt{simp05}; \citealt{can11}).

\subsection{Individual Calculations of \ion{H}{1} Hole Energetics}
\subsubsection{DDO 181}

DDO 181 was first listed in the dwarf galaxy catalog of \citet{vdb59}. Using the tip of the red giant branch, \citet{dal09}
estimate a distance to DDO 181 of 3.1 Mpc.  At this distance, the linear size of the \ion{H}{1} disk has a maximum of 
$\sim$4.4 kpc.  DDO 181 has an integrated \ion{H}{1} flux density of 11.5 Jy km s$^{-1}$ corresponding to a total \ion{H}{1} mass of
2.6$\times$10$^{7}$ M$_{\sun}$.  

The hole in DDO 181 is readily visible in the \ion{H}{1} integrated intensity map (Figure \ref{ddo181_hole}) as a
peanut shaped depression near the center of the \ion{H}{1} disk.  
To determine the size of the \ion{H}{1} hole we plot the average \ion{H}{1} column density versus radius (Figure \ref{rp}).  We defined the
center and ellipticity of the hole in DDO 181 visually from the \ion{H}{1} integrated intensity maps, placing 
the center of the hole at $\alpha = 13^{h}39^{m}52\fs1, \delta = 
+40{\degr}44{\arcmin}39{\farcs}0$.  Following the procedure outlined in \S\ref{holedef} we derive a radius of 740 pc. 
Using Equation \ref{ehole} we derive a range of hole creation energies of 3.9$\times$10$^{52}$ -
5.3$\times$10$^{53}$ erg.

\subsubsection{Holmberg I}

Holmberg I is a dIrr galaxy discovered by \citet{hol50} in a photometric study of nearby galaxies.  \citet{dal09} fit the tip 
of the red giant branch to determine a distance to Holmberg I of 3.9 Mpc.  Its distance and apparent $B$-band magnitude of 13.4 
give it an absolute $B$-band magnitude of $-$14.5, placing it among the least luminous observed dIrr galaxies \citep{beg08}.  
  The \ion{H}{1} integrated flux density of 
40.1 Jy km s$^{-1}$ corresponds to a total \ion{H}{1} mass for Holmberg I of 1.4$\times10^{8}$ M$_{\sun}$. 
Figure \ref{hoi_hole} is the \ion{H}{1} integrated intensity map which shows the large hole in the ISM, roughly 45$\arcsec$ South of the
dynamical center of the galaxy \citep{ott01}.  The \ion{H}{1} hole also dominates the optical disk of the galaxy (Figure
\ref{hoi_hole}).  We present the observed characteristics of Holmberg I in Table \ref{genprops}.

We perform the same radial analysis as above to characterize the size of the hole.  We plot the azimuthally averaged \ion{H}{1} column
density versus radius in Figure \ref{rp}.  The peak in this distribution is at $\sim$53$\arcsec$ or a radius of 1000 pc. 
\citet{ott01} derive a hole radius of $\sim$52$\arcsec$ but used a distance of 3.6 Mpc, giving the hole a radius of 850
pc.  Also, the VLA data has been reprocessed by the THINGS \citep{walt08}
reduction pipeline, changing some of the observed parameters.  For example, \citet{ott01} observed an average velocity
dispersion of roughly 9.0 km s$^{-1}$ whereas we measure 7.9 km s$^{-1}$.  
Despite these differences, their hole creation value ($E_{hole}$ $\lesssim$1.2$\times$10$^{53}$ erg)
falls within the range of values we derive for 
Holmberg I of 6.7$\times$10$^{52}$ - 1.2$\times$10$^{54}$ erg.  On the other hand, \citet{bag11} used the P-V diagram instead of the 
column density radial profile we use to derive the size of this same hole.  They derive a very discrepant hole radius of only 67.5 pc 
and thus a hole creation value of 2.4 $\times$ 10$^{50}$ erg.

A check of the plausibility of our hole creation values comes from the numerical study of \citet{vor04b}.  They simulated 
the \ion{H}{1} morphology of Holmberg I with a 2-D hydrodynamics code (ZEUS-2D) and injected different energies over 30 Myr to
approximate multiple SNe from a single age cluster of varying masses.  The model that best matched the observed morphology
of Holmberg I injected the energy of $\sim300$ SNe of 10$^{51}$ erg each or a total input energy of $3\times10^{53}$ erg. 
This value falls within our estimated hole creation energies from above.  

\subsubsection{M81 Dwarf A}

M81 Dwarf A was first listed in a survey of Sculptor type dwarf galaxies by \citet{kar68}. Eleven years later \citet{lo79}
rediscovered M81 Dwarf A in an atomic hydrogen search around Holmberg II in the M81 group of galaxies.  Recent estimates
place M81 Dwarf A at a distance of 3.4 Mpc \citep{dal09}.  This sets the observed
angular size of the \ion{H}{1} disk of $\sim220\arcsec$ equal to a linear diameter of $\sim$3.6 kpc.   The integrated \ion{H}{1} flux density 
of 4.1 Jy km s$^{-1}$ implies a total \ion{H}{1} mass of $1.1\times10^{7}$ M$_{\sun}$.  M81 Dwarf A was the least 
massive galaxy observed in the THINGS sample and is near the lower mass end of observed dIrr galaxies.  Table 
\ref{genprops} lists the relevant properties of M81 Dwarf A.

The detailed study of 
\citet{sar83} investigated the overall \ion{H}{1} morphology of M81 Dwarf A, pointing out the dominant ring feature seen in the ISM.  
Figure \ref{m81da_hole} shows the VLA \ion{H}{1} integrated intensity map of M81 Dwarf A.  This image reveals the large hole seen in 
the ISM which encompasses most of the optical disk (see Figure \ref{m81da_hole}).  
The peak in the average \ion{H}{1} column density versus radius (Figure \ref{rp}) corresponds to a radius of $\sim$45$\arcsec$ or 
$\sim$745 pc.  From this radius and properties listed in Table \ref{genprops} we derive a range of hole creation energies of
3.5$\times$10$^{51}$ - 5.0$\times$10$^{53}$ erg.

\subsubsection{Sextans A}

Sextans A is a relatively close galaxy at a distance of 1.4 Mpc \citep{dal09}.  Given its distance, it has also been extensively
studied.  \citet{skil88} studied the \ion{H}{1} distribution of Sextans A noting the ring-like central depression. \citet{puc94} described
this depression as ``twisted and sheared".  These authors also
point out that the majority of the optical disk is contained within the central depression (see Figure \ref{sexa_hole}).  
\citet{doh97} and \citet{doh02} used HST observations of the resolved stars to derive a SFH of Sextans A, noting that the most 
recent star formation is occurring near the peaks in the \ion{H}{1}.  \citet{vand98} used ground based observations to confirm these
results and went on to show that the older stars (50 - 100 Myr old) are more centrally concentrated than the youngest stars. 
This seems to follow a propagating star formation hypothesis \citep{ger78} suggested to occur in a supernova and stellar wind
driven hole expansion \citep{puc94}.  This pattern, however, is not seen in the HST data of \citet{doh02}.

Although the central depression is comprised of multiple holes in our observation as discussed in \S\ref{holedef}, for uniformity we treat it here as one large hole.  
We define the center of the hole to be located at  $\alpha = 10^{h}11^{m}01\fs1, \delta = -04{\degr}41{\arcmin}37{\farcs}0$, just offset
from a small, higher density cloud of gas to the east.  The average column density versus radius plot (Figure \ref{rp}) give us a
hole radius of 850 pc.  Figure \ref{sexa_hole} shows a blue circle denoting the boundary of the hole.  This radius combined
with the galaxy properties given in Tables \ref{genprops} and \ref{holeen} gives us a range of hole formation energies of 
1.1$\times$10$^{53}$ - 6.4$\times$10$^{53}$ erg.

\subsubsection{UGC 8508}

First cataloged by \citet{voro62}, UGC 8508 has been studied in detail by \citet{mou86}.  These authors conclude UGC 8508
to be at a state between quiescence and bursting based upon its optical colors.  UGC 8508 is relatively close by at a
distance of 2.6 Mpc \citep{dal09}.  Its \ion{H}{1} angular size of roughly 110$\arcsec$$\times$50$\arcsec$ corresponds to a linear size 
of about 1400$\times$630 pc.  UGC 8508 has an \ion{H}{1} integrated intensity of 13.8 Jy km s$^{-1}$ giving a total \ion{H}{1} mass of
2.2$\times$10$^{7}$ M$_{\sun}$.

The \ion{H}{1} hole is clearly observed in the Southeastern part of the disk as an elliptical depression (Figure \ref{u8508_hole}).  We
define the center of the hole to be at $\alpha = 13^{h}30^{m}45\fs9, \delta = +54{\degr}54{\arcmin}33{\farcs}0$.  From the
radial profile (Figure \ref{rp}) we determine the radius of the hole to be 285 pc leading to a hole creation energy range of 
9.2$\times$10$^{50}$ - 2.2$\times$10$^{52}$ erg.  This is the smallest and least energetic hole in our sample, yet it still
occupies a large fraction of the optical extent of the galaxy (Figure \ref{u8508_hole}).

\subsection{Energy input from stars \label{staren}}

\begin{figure}
\includegraphics[width=168mm]{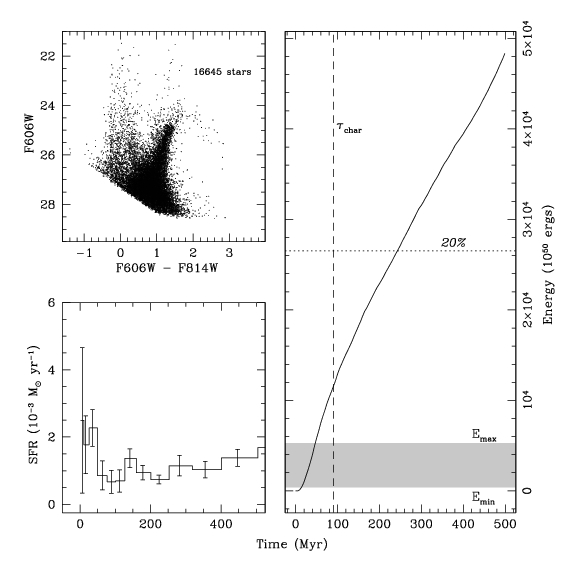}
\caption{DDO 181: $Upper~left$ - The (F606W, F606W-F814W) CMD for the stars within the the radius of the central \ion{H}{1} hole.
$Lower~left$ - The SFH over the past 500 Myr for the stars within the central \ion{H}{1} hole.  $Right$ - The cumulative energy due
to stellar winds and SNe calculated by using the SFH as input into STARBURST99 \citep{lei99}.  The shaded region denotes the
range in energy needed to create the \ion{H}{1} hole from section \ref{HIen}.  The vertical, dashed line is the characteristic age of the
hole.  The horizontal dashed lines represents the 10\% and/or 20\% feedback efficiency levels using the upper limit of hole creation energy. 
 \label{ddo181_sce}}
\end{figure}

\begin{figure}
\includegraphics[width=168mm]{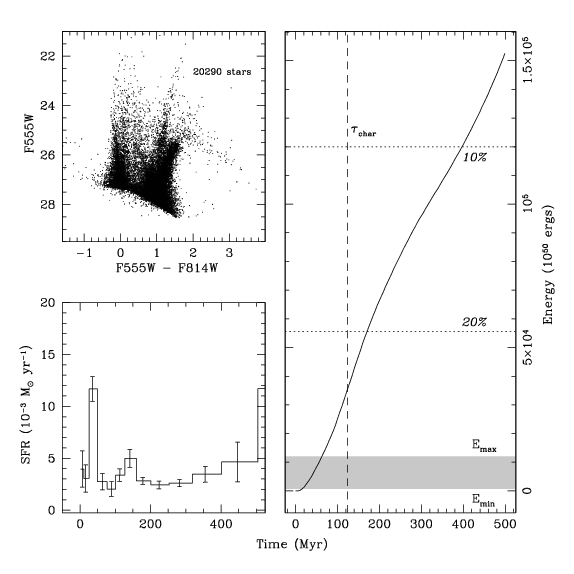}
\caption{Same as Figure \ref{ddo181_sce} except for Holmberg I. \label{hoi_sce}}
\end{figure}

\begin{figure}
\includegraphics[width=168mm]{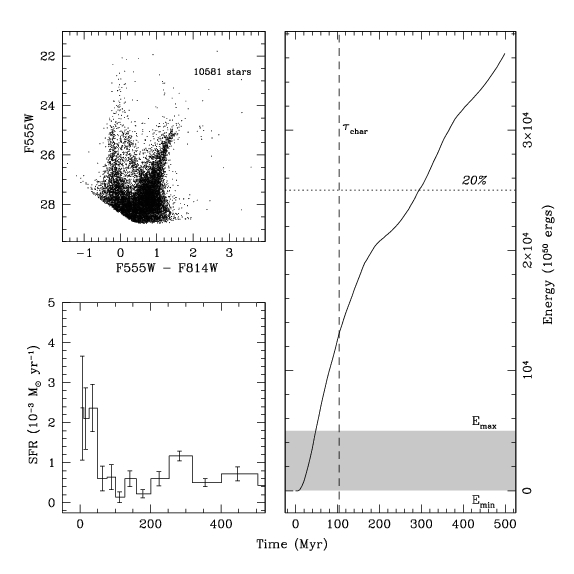}
\caption{Same as Figure \ref{ddo181_sce} except for M81 Dwarf A. \label{dwa_sce}}
\end{figure}

\begin{figure}
\includegraphics[width=168mm]{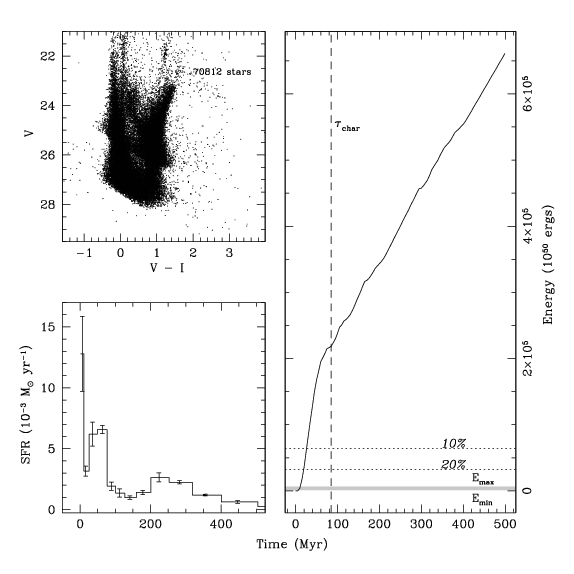}
\caption{Same as Figure \ref{ddo181_sce} except for Sextans A. The CMD is a combination of two different WFPC2 pointings of 
different integration times.
\label{sexa_sce}}
\end{figure}

\begin{figure}
\includegraphics[width=168mm]{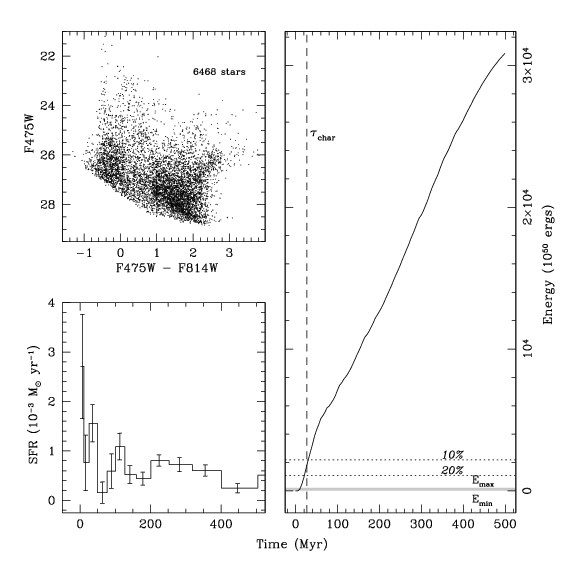}
\caption{Same as Figure \ref{ddo181_sce} except for UGC 8508. \label{u8508_sce}}
\end{figure}

To quantify the available energy from stellar processes we derive SFHs, i.e., a star formation rate as a function of time and metallicity,
from the resolved stars.  Color-magnitude diagrams (CMDs) contain the `fossil record' of 
star formation over the lifetime of a galaxy, from which we can determine a SFH.  The upper left panel of Figures 
\ref{ddo181_sce} - \ref{u8508_sce} show the CMDs of the stars contained within the holes.  We derived SFHs by the analysis of
HST-based optical CMDs using the technique described in 
\citet{dol02}.  Briefly, this method involves the construction of single age synthetic CMDs based on a specified stellar 
evolution model.  These individual synthetic CMDs are then linearly combined along with a model foreground CMD to produce 
a composite synthetic CMD.  Linear weights on the individual CMDs are adjusted to obtain the best fit as measured by a 
Poisson maximum likelihood statistic; the weights corresponding to the best fit represent the most probable SFH.

As input, we specified a standard power law initial mass function with x $=$ $-$2.3 from 0.1 to 100 M$_{\sun}$, a binary 
fraction $=$ 0.35, the stellar evolution models of \citet{mar08}, distances measured from the tip of the red giant branch 
\citep{dal09}, and foreground extinction values as specified by \citet{sch98}.  Each SFH calculation included the results from 
500,000 artificial 
star tests, in order to simulate observational errors and completeness.  For each galaxy we compared the completeness 
functions from the hole-only regions to the entire galaxy and found little difference.  This similarity is unsurprising as 
our sample of galaxies does not feature significant surface brightness gradients and the holes typically cover a large 
portion of the HST fields.  Because the CMDs considered in this analysis do not reach the ancient main sequence turnoff, the 
measured chemical enrichment is uncertain \citep[e.g.,][]{gal05}.  We thus constrained the 
metallicity of each SFH to monotonically increase as a function of time, preventing potentially unphysical chemical 
evolution models.

To quantify the errors in the SFHs, we consider both statistical and random uncertainties.  As a proxy for systematic 
effects, we computed the SFHs for a grid of small shifts ($\pm$0.05 mag) in distance and extinction.  The RMS per time bin from 
the grid of SFHs roughly probes systematic uncertainties in the stellar models.  However, \citet{wei11} show that this 
method may underestimate the magnitude of the systematic uncertainties by factors of a few.  For the purposes of the
energetics analysis in this paper, the adopted technique of deriving errors is sufficiently precise, given the larger 
uncertainties on HI energetic determinations as discussed in \S\ref{HIen}.  We further computed 50 Monte Carlo 
realizations of the model CMD to test for effects of Poisson fluctuations on the SFH, i.e., the random uncertainties.  The 
final error bars are simply the quadrature sum of the systematic and random errors.  
The lower left panel of Figures \ref{ddo181_sce} - \ref{u8508_sce} shows the derived SFHs over the past 500 Myr 
for the stars within the holes of each galaxy.  

Using the above methods we produce two sets of SFHs.  One set describes the global SFHs derived from all
of the stars within a given galaxy.  We can compare the global star formation histories from the galaxies in our sample that 
have created large \ion{H}{1} holes to SFHs of dwarf galaxies presented in the literature that have not created large \ion{H}{1} holes. 
We then look for distinguishing star forming features which may have created the observed structures.  
The second set of SFHs is derived using only the stars within the radius 
of the hole.  We use these SFHs to calculate the available energy from the evolving underlying stellar population. 

We next quantify the available energy from stellar evolution.
The amount of stellar feedback energy produced per time bin is calculated from the galaxy evolution modeling code 
STARBURST99 \citep{lei99}.  STARBURST99 uses stellar evolution models to estimate feedback (e.g., chemical,
spectrophotometric, and stellar energy evolution) for a given SFH.  We follow the method outlined in \citet{wei09} in which
we simulate a single burst of star formation with an initial mass of 10$^{6}$ M$_{\sun}$ and assuming a metallicity of 2\% Solar, 
which is consistent with the expected low metallicities of these low luminosity dwarf galaxies.   
We then normalize the output from the single
star formation simulation by the inferred mass from the star formation history, sampled in 5 Myr intervals.  
We plot the cumulative stellar input
energies on the right panel of Figures \ref{ddo181_sce} - \ref{u8508_sce} and discuss their implications in \S\ref{results}.

\section{Comparison with the Stellar Energy Budget \label{results}}

We now compare the energy available from stellar feedback processes from \S\ref{staren} to the range of estimated hole 
formation energies in Table \ref{holeen}.  The right panels of Figures \ref{ddo181_sce} - \ref{u8508_sce} show the cumulative
energy from stellar processes over the past 500 Myr.  This energy profile represents a {\it lower} limit to the available
stellar energy as energy estimates from STARBURST99 ignore any contribution of Type Ia supernovae (contained within the hole, but 
created by previous generations of star formation).  The vertical dashed line 
is the characteristic age ($\tau_{char}$) of the hole and the 
shaded region is the likely range of required hole formation energies (see \S\ref{HIen}).  In Table \ref{holeen} we list the total amount 
of energy created by the stars within the characteristic age (E$_{stars}$). In each galaxy E$_{stars}$ $\gg$ E$_{min}$ implying the 
amount of energy available due to stellar processes within the characteristic age of the holes exceeds the minimum energies for hole 
formation.  Using values of E$_{min}$ implies minimum stellar feedback efficiencies ($\epsilon_{min}$ = E$_{min}$/E$_{stars}$) of 
roughly 0.5\% to 4\%.

Alternatively, using the values for E$_{max}$ we derive maximum formation efficiencies ($\epsilon_{max}$ = E$_{max}$/E$_{stars}$) 
between 3\% and 48\% (see Table \ref{holeen}).  Note that the largest of these feedback efficiencies 
are above the range of $\sim$1\% - 20\% inferred by models of star formation interactions with the ISM  
(\citealt{the92}; \citealt{col94}; \citealt{pad97}; \citealt{tho98}).  However, our estimated characteristic ages are not particularly well
defined.  For example, \citet{wei09} provide
evidence that multiple generations of star formation are required to form and maintain large \ion{H}{1} holes.  \citet{rec06}
produced a model that maintains \ion{H}{1} holes for time scales as long as $\sim$600 Myr by injecting energy from multiple 
generations of SNe, prohibiting the gas from cooling and recollapsing. 

With these limitations in mind we can try to determine the age of the holes in another manner.  Since the models of interactions between
stellar feedback processes and the ISM predict efficiencies of 1\% - 20\%, we can use this to determine an age for the holes
assuming our upper limit for hole creation energy estimates to be correct.  
The dashed horizontal lines of Figures \ref{ddo181_sce} - \ref{u8508_sce} show efficiencies of 10\% and 20\% using the upper 
limit hole creation energy estimates.  For example, a 10\% feedback efficiency would require the energy of 10 times E$_{max}$ to create 
the hole.  These lines should give us upper limits on the ages of the holes for each feedback
efficiency.  Clearly the inferred characteristic ages are underestimating the likely age of formation of the holes in each of the galaxies
(except for UGC 8505) if energy input from multiple stellar populations have 
prohibited the holes from recollapsing.  The inferred characteristic age of the hole in UGC 8508 is comparable to the ages
determined for both 10\% and 20\% efficiencies, which is most likely due to the relatively small amount of required hole
creation energy. 

These results illustrate the significant level of uncertainty when dealing with large \ion{H}{1} holes that have blown out of the disk.  
Without the ability to accurately pin down the expansion velocity and initial distribution of the holes, estimating creation
energies and hole ages are highly uncertain and perhaps even arbitrary when one assumes the standard toy model of a single burst
of star formation.  More likely, multiple generations of star formation work in collaboration to grow these structures to
the sizes we observe today.  However, even with these uncertainties, it is clear that there is no evidence yet against forming these 
holes through stellar feedback.  Although the absolute values may vary, we have shown that the amount of energy available in the
underlying stellar population is more than enough to likely be the dominant creation source.

\section{Comparison of Global SFHs}

\begin{figure}
\includegraphics[width=168mm]{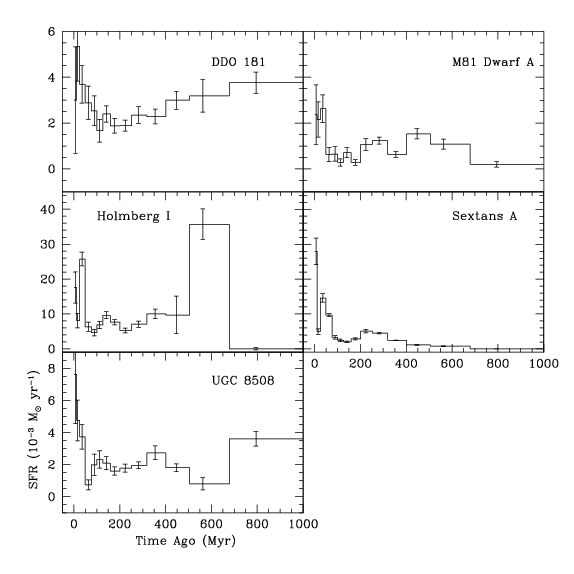}
\caption{Global SFHs derived using all of the resolved stars observed for each galaxy in our centrally dominant hole sample.  
There is no common signature of a single, dominant hole creation event in the SFHs when compared to each other. When compared
to the SFHs presented in \citet{doh98} and \citet{wei08}, the SFHs of this study do not distinguish themselves in any
way. \label{global}}
\end{figure}

We now address the question ``Is there an imprint on the global SFH of the event(s) that create the centrally
dominant hole in the ISM?".  To do so, we compare the recent {\it global} SFHs derived from all of the resolved stars of the five 
galaxies in this study to the recent global SFHs presented in \citet{doh98} and \citet{wei08}.  We present the recent 
global SFHs of the five galaxies in our sample in Figure \ref{global}.  There is no single feature beyond 100 Myr ago in common 
when comparing the five galaxies from our large hole sample with each other.  However, the most recent star formation activity ($<$100 Myr) 
for each galaxy in our sample shows a systematic rise that is likely due to star formation on the edges of our defined holes (see 
\S\ref{search}).  Likewise, when we compare the SFHs in our sample with those presented in
\citet{doh98} and \citet{wei08} nothing obvious
stands out as a major difference between the two populations. Many of the galaxies have a systematic rise in the most recent time bins 
whether they have created a large hole or not.  Our sample contains galaxies with very low, near constant levels of star
formation over the last Gyr and certainly provides no obvious recent burst that can be interpreted as a ``hole creation" signature
(e.g., GR8; \citealt{doh98} and M81 Dwarf A).  Our sample also includes galaxies that have significant star formation activity in their
recent histories.  However, some of these galaxies have created large, centrally dominant holes (i.e., Holmberg I), but others have 
not (i.e., NGC 2366; \citealt{wei08}).  This result, combined with the results from the previous section, suggests that sustained 
star formation is a necessary phenomenon to create large \ion{H}{1} holes, but not a sufficient one.  

This global comparison confirms the general
results of \citet{wei09}.  These authors looked at the many holes seen in the \ion{H}{1} distribution of Holmberg II and
compared the SFHs of the stars within the holes to those within control fields.  The control fields consisted of regions of
similar size but in locations where there were no \ion{H}{1} holes.  The expectation in this previous study was to find an 
increased level of star
formation in the regions of the holes compared to the control fields.  However, the control fields
contained amounts of energy comparable to (and in some cases greater than) the areas with holes.  This
surprising result suggests that the local conditions of the ISM and where stars form in the galaxy play a critical factor in whether or
not a hole is created.  For example, perhaps the surrounding ISM must be dense enough over a large region to sustain
a coherent structure prior to blow out.  It is likely that the porosity of the ISM plays a very important role.  A more 
porous medium may prohibit the formation of holes by allowing the ionized gases needed for pressure support to leak out 
before structures can be formed.  Also, the output energy produced by star formation occurring in multiple regions within a galaxy 
may destructively interfere, thus destroying newly formed holes.  In contrast, if star formation occurs in a consecutive and 
concentrated manner, the output energy may result in holes that combine to create the observed large features.

An alternative to the suggestion that these large holes are created by sustained star formation over multiple generations is that 
the stars play no role whatsoever and that local conditions of the ISM (e.g., gravitational instabilities) alone determine where 
holes are created.  We do not favor 
this explanation since it is known that stellar processes do indeed contribute to the heating and ionization of the ISM.  Our
results for the large amount of available stellar energy from \S\ref{results} and the results of the SFH comparison above seem to
confirm that the stars must play a role in creating the holes, however, not at the one-to-one level.  The local conditions of the
ISM must also play a critical role in this process.  However, the overall importance of each process is difficult to measure.

Numerical studies may be helpful to test which conditions are needed for stellar processes to create large 
\ion{H}{1} holes.  This exercise is beyond the scope of this work, but we can suggest some areas of
focus.  Ideally, 3-D magnetohydrodynamical simulations of a realistic dwarf galaxy ISM should be
modeled.  A parameter search should then follow including, but not necessarily limited to, simulating different
total \ion{H}{1} masses, varying ISM porosity values, and a range of star formation histories and locations within the disk.  These
suggestions would expand upon the pioneering work of \citet{rec06} and \citet{stin07}.  \citet{rec06} showed that multiple
generations of star formation can support large \ion{H}{1} structures for $\sim$500 Myr or more.  \citet{stin07} suggested that
dwarf galaxies undergo rapid star formation in which the gas is pushed out of the galaxy core, cools, and infalls back to the core
where the process is repeated in a pulsation-like ``breathing mode" on 100 Myr timescales.  At first glance, our findings seem to
support the longer duration hole ages of \citet{rec06}, and not the short-lived holes with a ``breathing mode" as suggested by
\citet{stin07}.  We can not rule out the model of \citet{stin07} on longer breathing mode time scales, however.  
At a minimum, new models would give insight into which, if any, of these parameters plays the biggest role in the creation 
and sustainability of large \ion{H}{1} holes.

\section{Searching for a Progenitor \label{search}}

\begin{figure}
\begin{center}
\includegraphics[width=110mm]{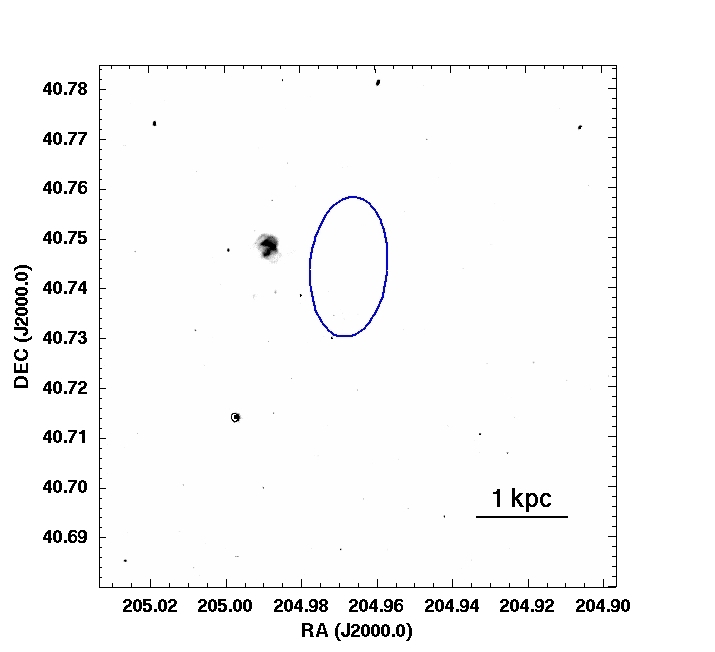} \\
\includegraphics[width=110mm]{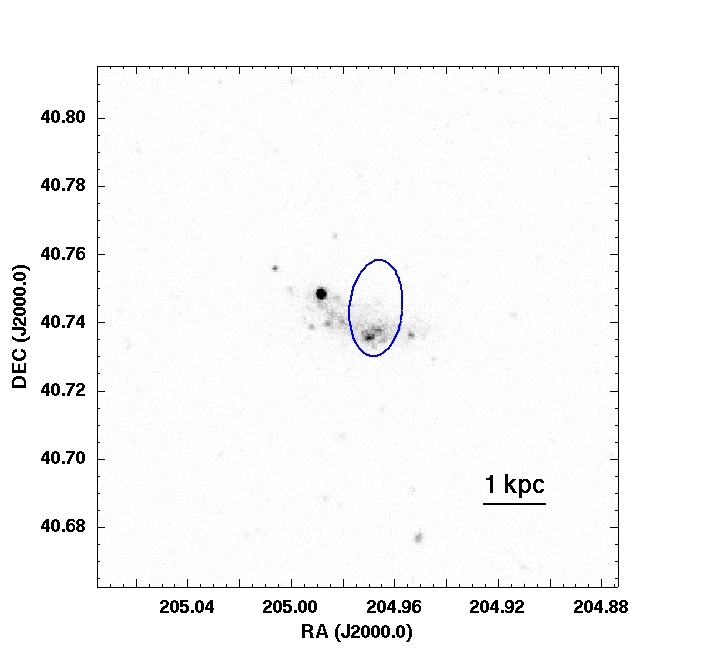} \\
\end{center}
\caption{Ground based H$\alpha$ (top) and GALEX FUV (bottom) image of DDO 181.
The blue ellipses denote the area of the hole.
\label{d181_prog}}
\end{figure}

\begin{figure}
\begin{center}
\includegraphics[width=110mm]{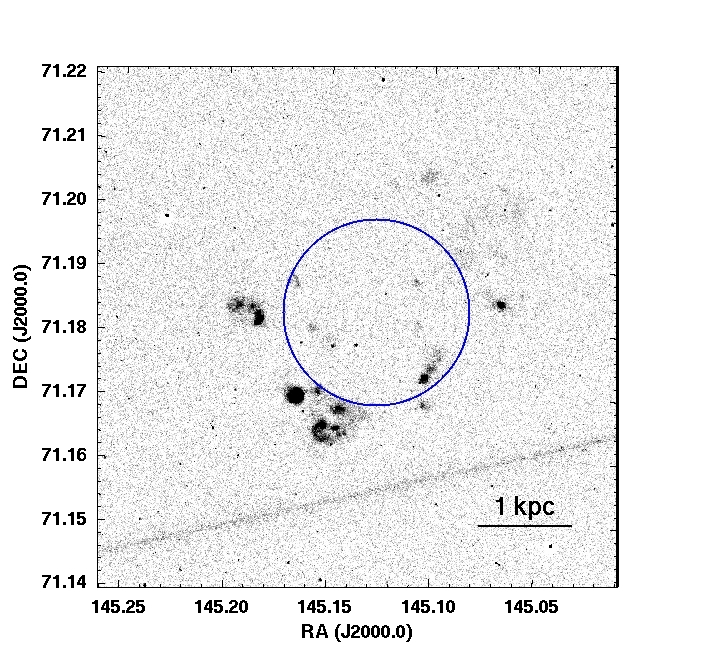} \\
\includegraphics[width=110mm]{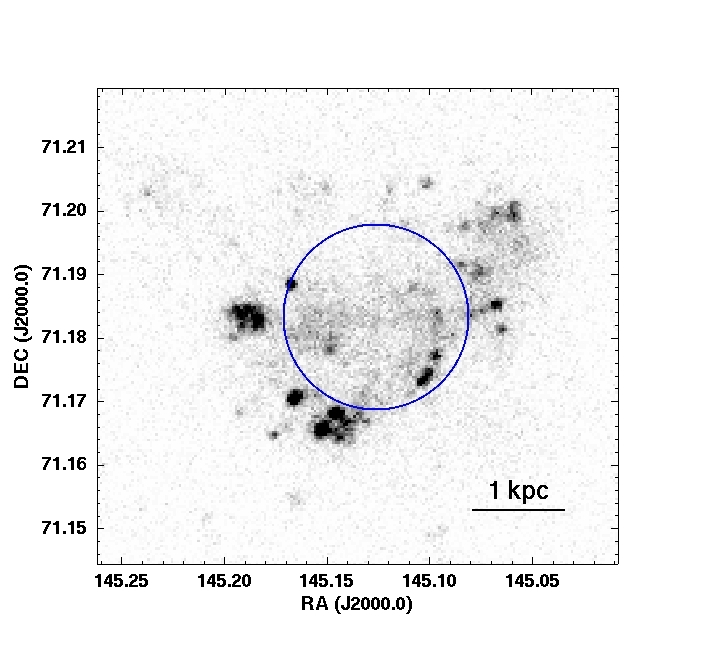} \\
\end{center}
\caption{Ground based H$\alpha$ (top) and GALEX FUV (bottom) image of Holmberg I. The blue circles denote the area of the hole.
\label{hoi_prog}}
\end{figure}

\begin{figure}
\begin{center}
\includegraphics[width=110mm]{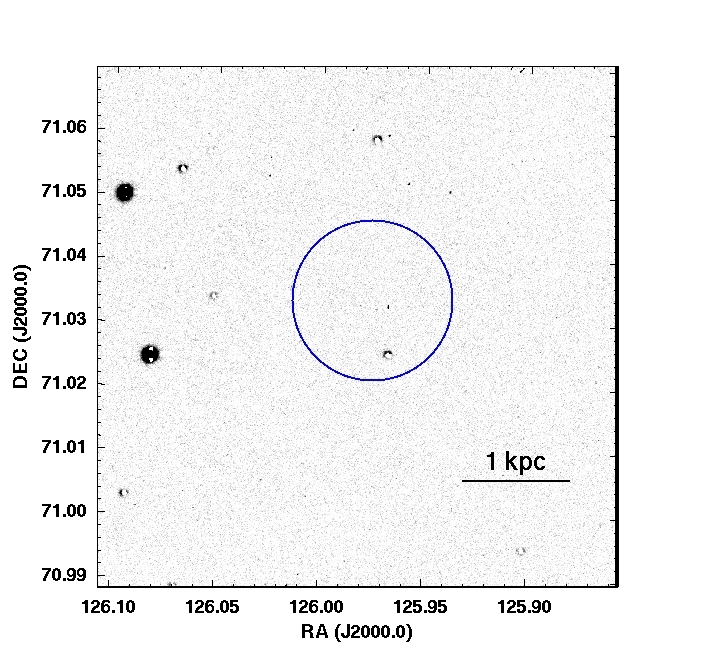} \\
\includegraphics[width=110mm]{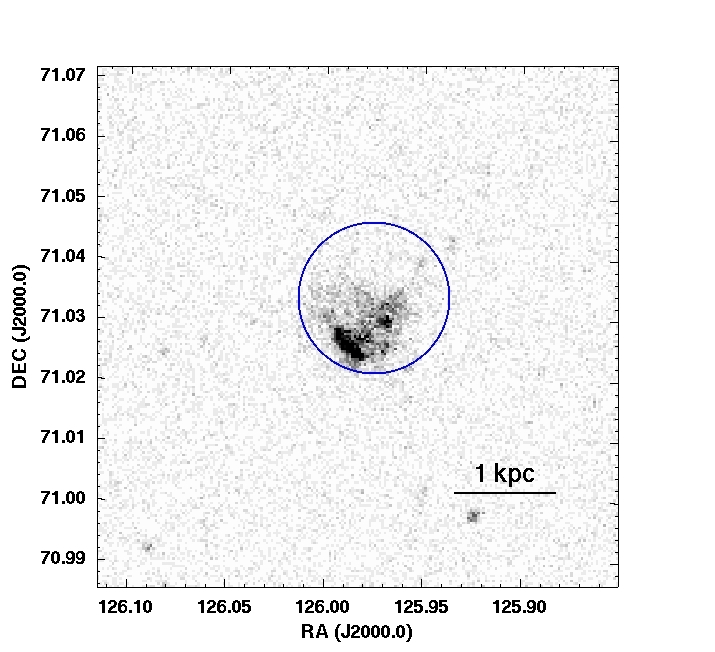} \\
\end{center}
\caption{Ground based H$\alpha$ (top) and GALEX FUV (bottom) image of M81 Dwarf A. The blue circles denote the area of the hole.
\label{mda_prog}}
\end{figure}

\begin{figure}
\begin{center}
\includegraphics[width=110mm]{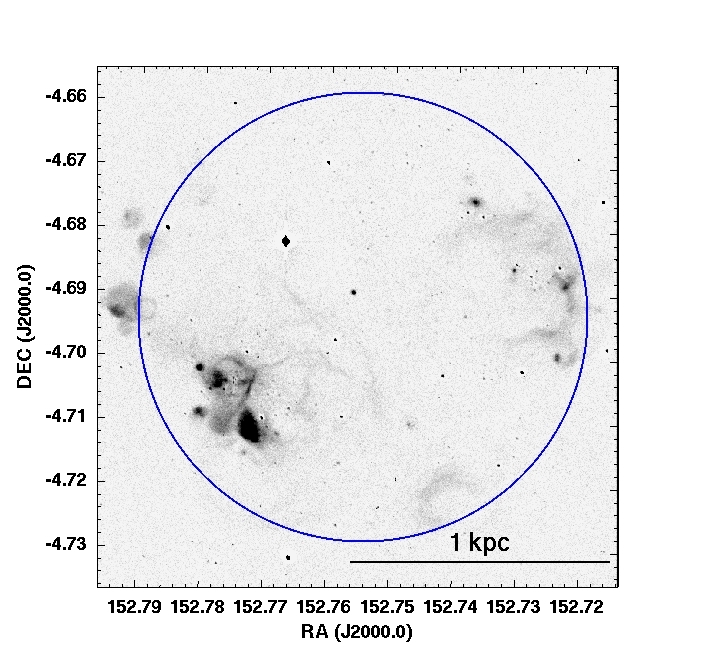} \\
\includegraphics[width=110mm]{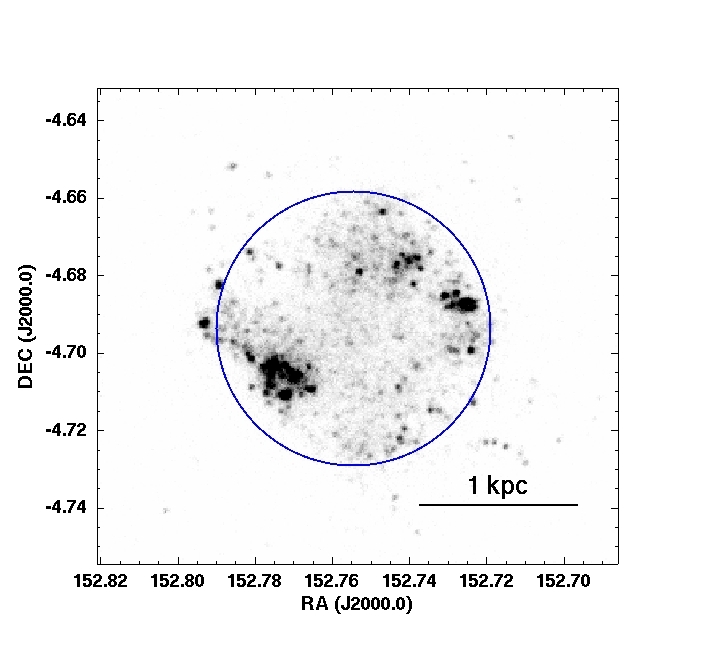} \\
\end{center}
\caption{Ground based H$\alpha$ (top) and GALEX FUV (bottom) image of Sextans A. The blue circles denote the area of the hole.
\label{sexa_prog}}
\end{figure}

\begin{figure}
\begin{center}
\includegraphics[width=145mm]{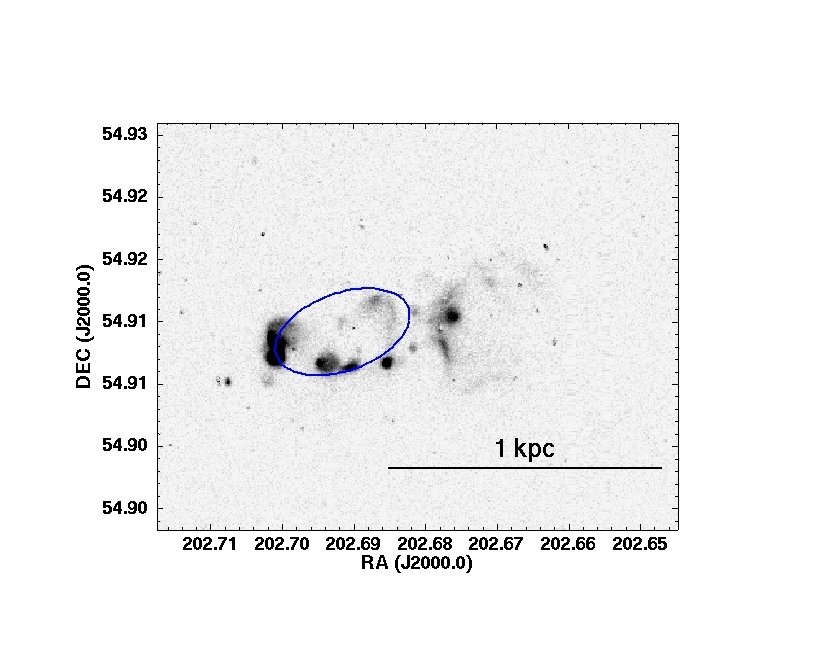} \\
\end{center}
\caption{Ground based H$\alpha$ image of UGC 8508. The blue ellipse denotes the area of the hole.
\label{u8508_prog}}
\end{figure}

If the large hole in each of our galaxies were formed by a single, centralized stellar association and not via multiple stellar
generations as we suggest, then observational evidence for the remnant cluster should exist.  To try and identify these clusters we 
use GALEX FUV and ground based H$\alpha$ imaging.  Each of these wavelengths probe stellar life cycles over different timescales.  UV 
wavelengths probe timescales up to $\sim$100 Myr while H$\alpha$ probes shorter $\sim$10 Myr timescales.  Note that these timescales
are shorter than the $\sim$200 Myr stellar dissolution timescales seen by \citet{bast11}.  Figures 
\ref{d181_prog} - \ref{u8508_prog}
show the FUV and H$\alpha$ images for each of our galaxies.  We have overlaid the hole location onto each 
image for comparison.  UGC 8508 could not be observed by GALEX due to bright foreground stars 
and/or too bright background levels \citep{lee11} so we only show the H$\alpha$ image.

The FUV and H$\alpha$ images show diffuse emission within some of the holes in our sample but no bright, central feature expected 
for a cluster.  However, each galaxy has bright features associated with recent star formation at or near the edges of our defined 
hole. DDO 181 has a bright FUV feature towards the southern edge of the the hole that has no associated H$\alpha$ emission.  
In Holmberg I, FUV and associated H$\alpha$ emission features are located around most of the edge of the hole.  The southern half of M81 
Dwarf A has FUV emission, with most of it on the southeastern edge of the hole; however,  M81 Dwarf A was not detected in H$\alpha$.
The large hole in Sextans A is filled with faint, diffuse and point-like FUV and H$\alpha$ emission.  The brightest features are 
near the southeastern and northwestern edges of the hole.  Lastly, UGC 8508 has bright H$\alpha$ knots all around the edge of the 
defined hole.  Each of the above bright FUV and H$\alpha$ features are associated with the densest \ion{H}{1} columns, and are not 
located in the center of the hole.  This alone does not mean that multiple generations of stellar processes are responsible for the
formation of these holes, but it does seem to rule out a young, single-aged cluster as being responsible.  

\citet{ten88} point out that 
large holes may have a stellar age gradient where the oldest stars
populate the inner regions and the youngest stars fall along the edges of the hole.  In this scenario of self-propagating star
formation we would expect to see knots of H$\alpha$ and FUV emission at or near the edge of the observed hole where the
conditions of star formation are suitable.  However, we do not expect a one-to-one relationship between the hole boundaries and 
the H$\alpha$ and FUV emission.  This is due to the possibility that some of the current star formation is not associated with the 
large hole (i.e., stochastic star formation).  None-the-less, we indeed see hints of possible propagating star formation in some 
of the Ha and FUV images.  These observations are consistent with the hypothesis that multiple generations of stars may provide 
the mechanical energy to create the large hole in each galaxy.  

Sextans A is perhaps the strongest case for multiple generations of star formation creating large holes in the ISM of these galaxies.  
Sextans A's relatively close distance of 1.4 Mpc allows us to resolve ISM structures on scales of $\sim$75 pc (compared to the $\sim$300 
pc reported by \citealt{skil88}).  The central depression breaks up into multiple smaller holes when observed at our high angular resolution.  
Perhaps some, if not all, of the large holes in the other galaxies in our sample would also break up into multiple, smaller holes at higher
angular resolution and signal to noise. Even with our incorrect assumption of a very large, single hole in Sextans A, the underlying stellar population has 
produced an ample amount of energy to have created a single, large hole, much less many smaller holes.  The H$\alpha$ and FUV emission only 
cement the idea that multiple generations of star formation, not single-aged clusters, produce large \ion{H}{1} holes.  

\section{Conclusions}

We have investigated the genesis of large \ion{H}{1} holes in the ISM of five nearby dwarf irregular galaxies by comparing the
energy required to create the holes to the inferred input energy from the underlying stellar populations.  In each galaxy, we have
shown that the input energy from the stars far exceeded the required energy derived from a single blast wave
model.  However, based on the observations here, there is no evidence of a single star formation 
event associated with hole creation.  Considering there is ample energy in the underlying stellar population, we conclude that 
multiple generations of stars are likely responsible for the creation and support of these large structures.

However, since there seems to be no correlation between the SFHs and a resultant hole, stars can not be the only player in this 
game.  It seems plausible that the local conditions of the
ISM must also be suitable for \ion{H}{1} holes to be created and sustained.  Star formation has been shown to occur on the rims of
holes (e.g., \citealt{wei09}), and indeed some of the UV and H$\alpha$ images of our sample galaxies also show evidence of star formation 
on the rim of the large \ion{H}{1} holes.  These new sites of star formation can either help inflate the holes to larger sizes or destroy the 
holes by pushing the gas back into the cavities.  It is presumably this delicate interplay between these effects that will ultimately
determine whether a galaxy will or will not form a large \ion{H}{1} hole.

In summary, the idea that a single-aged cluster is required to form the holes does not work based on the evidence at hand.  Past
numerical studies of hole creation have relied upon this tenet (see, for example,
\citealt{vor04b}), although some numerical studies intentionally use massive, single-aged clusters to produce Type 1 holes (e.g., \citealt{mac98}).  
Future numerical studies aiming to accurately investigate hole creation should
consider focusing on multiple generations of star formation occurring at different locations throughout the disk.  This study
reaffirms that the physics of the ISM of dwarf irregular galaxies is complicated and one can not use
simplified approximations to predict observed distributions. 

\acknowledgments{We thank the anonymous referee for a prompt and detailed report which significantly improved 
the clarity of the manuscript.  
Support for this work was provided by NRAO through the
National Science Foundation collaborative research grant 807515.  
NRAO is operated by Associated Universities,
Inc., under cooperative agreement with the National
Science Foundation. SRW is grateful for support
from a Penrose Fellowship.  This research
has made use of NASA's Astrophysics Data System Bibliographic
Services and the NASA/IPAC Extragalactic
Database (NED), which is operated by the Jet Propulsion
Laboratory, California Institute of Technology, under 
contract with the National Aeronautics and Space Administration.}

\clearpage

\begin{table}
\begin{center}
\caption{Beam and Resolution of \ion{H}{1} Images \label{imageprops}}
\begin{tabular}{lcccc}
\hline\hline
Name & B$_{maj}$ & B$_{min}$& BPA & Velocity Resolution  \\
     & (\arcsec) & (\arcsec) & ($^\mathrm{o}$) & (km s$^{-1}$) \\
\hline
DDO 181     & 12.5 & 10.5 & -80.4 & 0.63 \\
Holmberg I  & 14.7 & 12.7 & -41.6 & 2.5  \\
M81 Dwarf A & 15.9 & 14.2 &  10.2 & 2.5  \\
Sextans A   & 11.6 & 10.9 &  47.6 & 0.63 \\
UGC 8508    & 14.0 & 11.5 &  88.1 & 0.63 \\
\hline
\end{tabular}
\end{center}
\end{table}

\clearpage

\begin{table}
\begin{center}
\tiny
\caption{General Galaxy Properties \label{genprops}}
\begin{tabular}{lccccccccccc}
\hline\hline
Galaxy & RA & DEC & d & scale & m$_{B}^{a}$ & M$_{B}$ & L$_{B}$ & M$_{HI}$ & N$_{HI,Peak}$ & N$_{HI,Ave}$ & $<\sigma_{v}>$ \\
 & (J2000.0) & (J2000.0) & (Mpc) & (pc/\arcsec) & & & (10$^{7}$ L$_{\sun})$ & (10$^{7}$ M$_{\sun})$ & (10$^{21}$ cm$^{-2}$) & (10$^{20}$ cm$^{-2}$) & (km s$^{-1}$) \\
\hline
DDO 181     & $13^{h}39^{m}53\fs8$ & $+40{\degr}44{\arcmin}21{\arcsec}$ & 3.1 & 15.0 & 14.4 & -13.0 & 1.12 & 2.60 & 1.77 & 3.12 & 8.2 \\
Holmberg I  & $09^{h}40^{m}32\fs3$ & $+71{\degr}10{\arcmin}56{\arcsec}$ & 3.9 & 18.9 & 13.4 & -14.5 & 4.45 & 14.6 & 2.26 & 2.78 & 7.9 \\
M81 Dwarf A & $08^{h}23^{m}55\fs1$ & $+71{\degr}01{\arcmin}56{\arcsec}$ & 3.4 & 16.5 & 16.3 & -11.4 & 0.26 & 1.07 & 0.61 & 1.09 & 7.0 \\
Sextans A   & $10^{h}11^{m}00\fs8$ & $-04{\degr}41{\arcmin}34{\arcsec}$ & 1.4 & 6.79 & 11.7 & -14.0 & 2.81 & 6.80 & 5.98 & 5.16 & 9.8 \\
UGC 8508    & $13^{h}30^{m}44\fs4$ & $+54{\degr}54{\arcmin}36{\arcsec}$ & 2.6 & 12.6 & 14.0 & -13.1 & 1.22 & 2.20 & 2.98 & 2.42 & 10.3\\
\hline
\end{tabular}
\end{center}
Columns are: galaxy name, Right Ascension (RA), declination (DEC), distance (d), image scale, apparent $B$-band magnitude (m$_{B}$), absolute blue
magnitude (M$_{B}$), blue luminosity (L$_{B}$), \ion{H}{1} mass (M$_{HI}$), peak and average \ion{H}{1} column density (N$_{HI,Peak}$, 
N$_{HI,Ave}$), and the average velocity dispersion measured in the second moment map ($<\sigma_{v}>$). \\
$^{a}$\citet{kar04}, apparent magnitudes are corrected for Galactic foreground extinction. \\

\end{table}

\clearpage

\begin{table}
\begin{center}
\tiny
\caption{Hole Properties \label{holeen}}
\begin{tabular}{lccccccccccc}
\hline\hline
Galaxy & RA & DEC & r$_{hole}$ & P.A. & $\tau_{char}$ & E$_{stars}$ & E$_{max}$ & E$_{min}$ & $\epsilon_{max}$ & $\epsilon_{min}$ & age$_{alt}$ \\
& (J2000) & (J2000) & (pc) & ($\degr$) & (Myr) & (ergs) & (ergs) & (ergs) & & & (Myr) \\
\hline
DDO 181     & $13^{h}39^{m}52\fs1$ & $+40{\degr}44{\arcmin}39{\farcs}0$ & 755 (415)  & 85 & 90  & 1.1$\times$10$^{54}$ & 5.3$\times$10$^{53}$ & 3.9$\times$10$^{52}$ & 48\% & 3.5\% & $>$500, 240 \\
Holmberg I  & $09^{h}40^{m}30\fs$  & $+71{\degr}11{\arcmin}01{\farcs}8$ & 1000       & 0  & 124 & 3.4$\times$10$^{54}$ & 1.2$\times$10$^{54}$ & 6.7$\times$10$^{52}$ & 35\% & 2.0\% &    395, 170 \\
M81 Dwarf A & $08^{h}23^{m}54\fs1$ & $+71{\degr}02{\arcmin}01{\farcs}5$ & 745        & 0  & 104 & 1.4$\times$10$^{54}$ & 5.0$\times$10$^{53}$ & 3.5$\times$10$^{51}$ & 35\% & 0.3\% & $>$500, 390 \\
Sextans A   & $10^{h}11^{m}01\fs1$ & $-04{\degr}41{\arcmin}37{\farcs}0$ & 850        & 0  & 85  & 2.2$\times$10$^{55}$ & 6.4$\times$10$^{53}$ & 1.1$\times$10$^{53}$ & 3\%  & 0.5\% &     27, 19  \\
UGC 8508    & $13^{h}30^{m}45\fs9$ & $+54{\degr}54{\arcmin}33{\farcs}0$ & 285 (157)  & 20 & 27  & 1.8$\times$10$^{53}$ & 2.2$\times$10$^{52}$ & 9.2$\times$10$^{50}$ & 12\% & 0.5\% &     30, 20  \\
\hline
\end{tabular}
\end{center}
Columns are: galaxy name, Right Ascension (RA) and declination (DEC) of the center of the hole, the hole radius (semi-major axis
length for DDO 181 and UGC 8508; semi-minor axis length is in parentheses), the position angle of the hole (P.A.),
the characteristic age of the hole
($\tau_{char}$), the cumulative stellar energy budget at the characteristic age of the hole (E$_{stars}$), the upper limit energy for creating the hole
derived from Equation \ref{ehole} (E$_{max}$ is derived assuming N$_{HI}=10^{21}$ cm$^{-2}$), the lower energy limit to create the hole 
(E$_{min}$ is derived using the average observed N$_{HI}$), the upper and lower stellar feedback efficiencies ($\epsilon_{max}$ and $\epsilon_{min}$), 
and alternative ages (age$_{alt}$ are the age estimates for stellar feedback efficiencies of 10\% and 20\% using the upper limits of the hole creation energies 
(see \S\ref{results})). \\

\end{table}

\clearpage

\clearpage

\clearpage

\clearpage

\clearpage

\clearpage

\clearpage

\clearpage

\clearpage

\clearpage

\clearpage

\clearpage

\clearpage

\clearpage

\clearpage

\clearpage

\clearpage

\clearpage

\end{document}